\documentclass[aps,pre,reprint,floatfix,showpacs,groupedaddress]{revtex4-1}
\usepackage[T1]{fontenc}
\usepackage{url}
\usepackage[dvips]{graphicx}
\usepackage{amsmath,amsfonts,amssymb,amsbsy}  
\usepackage{psfrag,color,pstricks,pst-grad}

\newcommand{\bb}{\mathbf}
\newcommand{\ca}{\mathcal}

\newcommand{\hatb}{}
\newcommand{\bbf}{{}}

\begin{document}
\title{Thermal balance and quantum heat transport in nanostructures thermalized by local Langevin heat baths}
\author{K. S\"a\"askilahti}
\email{kimmo.saaskilahti@aalto.fi}
\author{J. Oksanen}
\author{J. Tulkki}
\affiliation{Department of Biomedical Engineering and Computational Science, Aalto University, FI-00076 AALTO, FINLAND}
\date{\today}

\begin{abstract}
 Modeling of thermal transport in practical nanostructures requires making trade-offs between the size of the system and the completeness of the model. We study quantum heat transfer in a self-consistent thermal bath setup consisting of two lead regions connected by a center region. Atoms both in the leads and in the center region are coupled to quantum Langevin heat baths that mimic the damping and dephasing of phonon waves by anharmonic scattering. This approach treats the leads and the center region on same footing and thereby allows for a simple and physically transparent thermalization of the system, enabling also perfect acoustic matching between the leads and the center region. Increasing the strength of the coupling reduces the mean free path of phonons and gradually shifts phonon transport from ballistic regime to diffusive regime. In the center region, the bath temperatures are determined self-consistently from the requirement of zero net energy exchange between the local heat bath and each atom. By solving the stochastic equations of motion in frequency space and averaging over noise \bbf{using the general fluctuation-dissipation relation derived by Dhar and Roy [J. Stat. Phys. \textbf{125}, 801 (2006)]}, we derive the formula for thermal current, which contains the Caroli formula for phonon transmission function and reduces to the Landauer-B\"uttiker formula in the limit of \bbf{vanishing coupling to local heat baths}. We prove that the bath temperatures measure local kinetic energy and can, therefore, be interpreted as true atomic temperatures. In a setup where phonon reflections are eliminated, Boltzmann transport equation under gray approximation with full phonon dispersion is shown to be equivalent to the self-consistent heat bath model. We also study thermal transport through two-dimensional constrictions in square lattice and graphene and discuss the differences between the exact solution and linear approximations. 
\end{abstract}
 \maketitle

\section{Introduction}

Recent theoretical and experimental studies of thermal properties of materials have demonstrated many exotic phononic phenomena such as ballistic and anomalous transport \cite{berber00,kim01,chang08}, conductance quantization \cite{rego98,schwab00}, and phonon tunneling \cite{prunnila10,altfeder10}. These discoveries suggest that the ability to manipulate heat flow at microscopic level and to better understand phonon transfer in nanoscale may lead to important technological breakthroughs ranging from new materials for thermoelectric conversion \cite{majumdar04,dubi11} to improved thermal management in future electronics \cite{pop10} and even information processing by phonons \cite{terraneo02, li06, chang06}.

Modeling of thermal transport in practical nanostructures typically requires making trade-offs between the size of the system and the completeness of the the model. Consequently, the commonly used models such as Boltzmann transport equation (BTE) \cite{murthy05}, molecular dynamics (MD), Landauer-B\"uttiker (LB) formalism \cite{landauer70,buttiker92} for phonon transfer \cite{rego98,mingo03} and full non-equilibrium Green's function (NEGF) method \cite{mingo06,wang06} each have distinct strengths and weaknesses. For instance, BTE for phonons is a powerful method that is applicable even for macroscopic systems, but it does not apply well to microscopic systems where wave effects such as diffraction are important.  MD can be applied to phonon transport in microscopic systems and accounts, e.g., for wave effects and phonon-phonon scattering due to anharmonicity of the interatomic potential, but it becomes computationally heavy for large systems and cannot strictly account for quantum statistics. The LB and NEGF models can fully account for the quantum statistics, but LB assumes ballistic phonon transfer and NEGF is computationally extremely demanding and therefore limited so far to very small systems \cite{wang08}.

As a consequence of the above limitations, none of the above models are well suited for modeling phonon transfer in typical nanostructures consisting of a relatively large number of atoms. A very interesting compromise between system size and model completeness is provided by the self-consistent thermal bath (SCTB) model suggested by Bolsterli, Rich and Visscher \cite{bolsterli70}. In the SCTB model, the phonon scattering is mimicked by coupling the atoms to local heat reservoirs whose temperatures are determined from the condition that, in the steady state, there is no net energy transfer between an atom and the corresponding local heat reservoir. The concept was first used to show that for a classical system with bath temperatures equal to the local kinetic temperatures the thermal conductivity of a harmonic one-dimensional chain was rendered finite by the bath couplings. Later it was shown rigorously that in an infinite one-dimensional chain in a non-equilibrium steady state, the system is at local thermal equilibrium \cite{bonetto04} and that local heat current is proportional to the thermal gradient, i.e. heat transfer is diffusive \cite{bonetto04,pereira04,dhar06,roy08}. SCTB model has also been applied to investigating quantum effects in non-ballistic heat transfer \cite{visscher75,dhar06,roy08,bandyopadhyay11}, effects of additional anharmonicity \cite{pereira04,falcao08,bonetto09,pereira13} and unequal masses \cite{barros06,neto07,santana12} on heat conduction and the necessary ingredients of thermal rectification \cite{pereira08,segal09,pereira10,pereira11,bandyopadhyay11,avila13}. 

Self-consistent heat baths are closely related to B\"uttiker's self-consistent voltage probes \cite{buttiker85,buttiker86}, which are employed in electron transport as models for dephasing and dissipation caused by inelastic scattering. To account for the inelastic effects using a microscopic model for the voltage probe, D'Amato and Pastawski modeled the probes by one-dimensional tight-binding chains \cite{damato90} and were able to demonstrate a transition from coherent to diffusive transport. Their work was recently extended by Roy and Dhar \cite{roy07} to cover simultaneous charge and heat transfer in the presence of a chemical potential and temperature gradient. Momentum-conserving scatterers have also been proposed \cite{golizadeh07}. 

In this paper, we extend the SCTB models beyond one-dimensional chains and study the heat transfer and the use of SCTB models in describing quantum thermal transfer in one-dimensional and two-dimensional structures that exhibit geometric as well as phonon-phonon scattering. To describe the dissipative effects in the whole infinite system consisting of two leads and the center region, atoms both in the leads and in the self-consistent center region are coupled to Langevin heat baths. This makes our setup different from the situation considered by Dhar and Shastry in Ref. \cite{dhar03} and Dhar and Roy in Ref. \cite{dhar06}, describing purely ballistic phonon transport in the leads. We compare the predictions of SCTB with Landauer-B\"uttiker (LB) formalism and Boltzmann transport equation (BTE). In contrast to LB formalism, phonon transport in the SCTB model is not purely ballistic due to the interaction with the local heat baths, but we show that the SCTB model reduces to the conventional LB model in the limit of \bbf{vanishing coupling to local heat baths}. In a setup where wave effects can be neglected, SCTB is shown to be equivalent to BTE under gray approximation. We demonstrate how the local bath temperatures are intuitively related to the local energy densities. We compare the exact self-consistent temperature profiles to linear and classical approximations and thereby extend the work by Bandyopadhyay and Segal \cite{bandyopadhyay11}, who, in contrast to the semi-infinite leads studied here, considered purely Ohmic lead couplings. We also extend their work on one-dimensional chains by comparing the quantum and classical temperature profiles in higher-dimensional structures.

The paper is organized as follows. In Sec. \ref{sec:theory}, we present the computational setup and derive the formula for heat currents flowing to the leads and to self-consistent heat baths. This is achieved by first solving the Heisenberg-Langevin equations of motion in Sec. \ref{sec:equations_of_motion} and then specifying the statistical properties of noise terms in Sec. \ref{sec:noise_properties}. Sections \ref{sec:heatflowtobaths} and \ref{sec:temperature_properties} are devoted to calculating the average heat flow to the baths and presenting physical interpretation for the self-consistent bath temperatures. For comparison purposes, we also solve the Boltzmann transport equation under gray approximation for the one-dimensional chain in Sec. \ref{sec:bte_theory}.

In Sec. \ref{sec:methods}, we discuss how to solve the self-consistent equations either by iterative means or by linearization. We also present a physically intuitive method of solving the equations, which can be interpreted as describing the transient behavior of the system. As an application of the formalism, we study in Sec. \ref{sec:rg_chain} heat transfer in a one-dimensional chain coupled to semi-infinite chains, in the so-called Rubin-Greer setup \cite{rubin71}, and highlight the connection to Boltzmann transport equation. We then study thermal transport through two-dimensional constrictions both for square lattice and the more practical case of graphene in Secs. \ref{sec:2d_junction} and \ref{sec:graphene}. The methods for solving the self-consistent non-linear equations are compared in Sec. \ref{sec:results_comparison}. Conclusions are finally given in Sec. \ref{sec:conclusions}.

\section{Theory}
\label{sec:theory}
\begin{figure}
 \includegraphics[width=8.6cm]{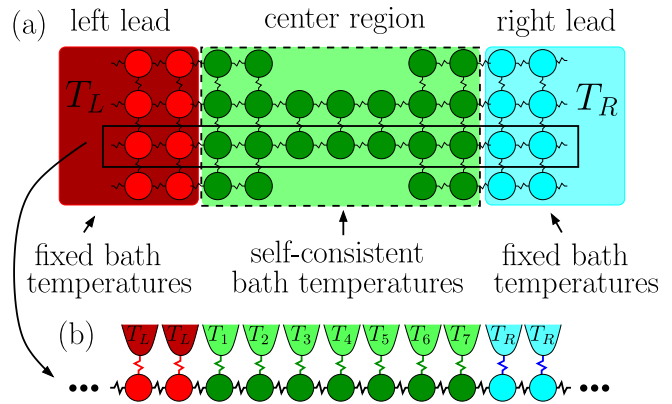}
 \caption{(Color online) (a) A schematic illustration of the system under study. The structure is divided into the left lead, the center region and the right lead. All atoms are coupled to spatially uncorrelated quantum Langevin heat baths, which are shown explicitly for one cross-section in (b). In the left and right lead, the temperatures of the local heat baths have prescribed values $T_L$ and $T_R$, respectively. In the center region, on the other hand, temperature varies between $T_L$ and $T_R$ and the bath temperatures are determined self-consistently using the requirement that the average thermal current to each bath vanishes. The leads can contain an infinite number of atoms, but the center region is finite. Two-dimensional square lattice with nearest neighbor interactions is shown for illustrative purposes, but the basic principle can be applied to any geometry.}
\label{fig:sud1}
\end{figure}

In the theory and results of this paper, we mainly focus on a system that essentially consists of the left lead region, center region and right lead region as shown schematically in Fig. \ref{fig:sud1}. All atoms within the leads are coupled to local Langevin heat baths set to prescribed values $T_L$ and $T_R$. The atoms in the center region are coupled to local heat baths whose temperatures are determined self-consistently from the requirement of local current conservation. The coupling to the Langevin heat baths effectively mimics thermalizing events such as phonon-phonon scattering. It is important to stress that in contrast to Landauer-B\"uttiker model, phonon transport is \textit{not} assumed to be ballistic either in the leads or in the center region. Although our approach of integrating out the leads, detailed below, is inspired by the work of Dhar and Shastry \cite{dhar03} and Dhar and Roy \cite{dhar06}, the thermalization in the leads in our setup takes place through a coupling to heat baths instead of thermalization by Ford-Kac-Mazur formalism \cite{ford65}. This method is physically transparent, since no difference is made between the leads and the center region (except for the bath temperatures). The method also allows to include dissipative effects in the leads, thereby enabling perfect acoustic matching between the leads and the center region.

In the following, we first solve the equations of motion in Sec. \ref{sec:equations_of_motion}, then specify the statistical properties of the noise in Sec. \ref{sec:noise_properties} and finally derive the formula for heat currents in Sec. \ref{sec:heatflowtobaths}.

\subsection{Equations of motion}
\label{sec:equations_of_motion}

The time evolution of atoms in the setup of Fig. \ref{fig:sud1} consists of two parts. The first part is deterministic and is specified by the system Hamiltonian $\ca{H}$ and Heisenberg equations of motion. The second part consists of a stochastic force and friction due to the interaction with the local heat bath and cannot be directly derived from a Hamiltonian \cite{weiss}.

The Hamiltonian time evolution of the atomic displacement $u_i^{\alpha}$ of atom $i$ along direction $\alpha\in\{x,y,z\}$ and corresponding conjugate momentum $p_i^{\alpha}$ is determined by the Hamiltonian $\mathcal{H}$ and the Heisenberg equations of motion $\dot{u}_i^{\alpha}=(i/\hbar)[\ca{H},u_i^{\alpha}]$ and $\dot{p}_i^{\alpha}=(i/\hbar)[\ca{H},p_i^{\alpha}]$. Here $[A,B]=AB-BA$ denotes the commutator and the atomic displacement $u_i^{\alpha}=q_i^{\alpha}-q_i^{\alpha 0}$ is defined as the variation of position $q_i^{\alpha}$ from the equilibrium position $q^{\alpha 0}_i$. The Hamiltonian of the system is, in the harmonic approximation,
\begin{equation}
 \ca{H} = \frac{1}{2} \sum_{I} \left[ \frac{\bb{p}_I^2}{m} + \bb{u}_I^T \bb{K}_I \bb{u}_I \right] +  \frac{1}{2} \sum_{I} \sum_{J\neq I} \bb{u}_I^T \bb{V}_{IJ} \bb{u}_J,
 \label{eq:Hfull_def}
\end{equation}
where index $I\in\{C,L,R\}$ labels the region: $C$ stands for center region, and $L$ and $R$ for the left and right leads, respectively. The displacement and momentum vectors $\bb{u}_I$ and $\bb{p}_I$ contain the displacements and momenta of all particles in region $I$ and we assume the masses $m$ of all atoms to be equal for notational simplicity. The spring constant matrix $\bb{K}_I$ and the coupling matrices $\bb{V}_{IJ}$ are the block components of the full spring constant matrix $\bb{K}$ divided into blocks as
\begin{equation}
 \bb{K} =\left( \begin{matrix}
           \bb{K}_L & \bb{V}_{LC} & 0\\
	   \bb{V}_{CL} & \bb{K}_C & \bb{V}_{CR} \\
	   0	& \bb{V}_{RC} & \bb{K}_R \\
          \end{matrix} \right),
\end{equation}
where we assumed that the leads do not interact, so $\bb{V}_{LR}=\bb{V}_{RL}^T=0$. The elements of $\bb{K}$ are obtained from the second derivative of the interatomic interaction energy $\ca{V}$ as \cite{ziman} 
\begin{equation}
 K_{ij}^{\alpha\beta} = \left. \frac{\partial^2 \ca{V}}{\partial q_i^{\alpha} \partial q_j^{\beta}}\right|_{\bb{q}=\bb{q}^0}.
\end{equation}
The equilibrium positions are defined by the condition of zero force
\begin{equation}
 \left.\frac{\partial \ca{V}}{\partial q_i^{\alpha}}\right|_{\bb{q}=\bb{q}^0}=0 ,
\end{equation}
which must be satisfied for all atoms $i$ and components $\alpha$.

The Heisenberg equations of motion that follow from the quadratic Hamiltonian \eqref{eq:Hfull_def} coincide with the classical equations of motion. Accompanied with the non-Hamiltonian time-evolution arising from the interaction with the heat bath, the equations of motion become
\begin{equation}
  m\ddot{\bb{u}}_I = - \bb{K}_I \bb{u}_I - \sum_{J\neq I} \bb{V}_{IJ} \bb{u}_J - m \gamma_I \dot{\bb{u}}_I + \xi_I. \label{eq:eom_I}
\end{equation}
The last two terms are Langevin friction and noise terms that turn the Heisenberg equation of motion into a quantum Langevin equation \cite{ford88,dhar06,weiss}. The stochastic force $\xi_I$ is a vector whose $i$'th component is the fluctuating force at site $i$ due to the interaction with the local heat bath. The statistical properties of the Langevin terms are discussed in the next section. 

Focusing on the steady-state behavior enables solving Eq. \eqref{eq:eom_I} by Fourier transformation defined, as usual, by $\hat f(\omega) = \int dt e^{i\omega t} f(t)$ with the corresponding inverse transformation $ f(t) = \int (d\omega/2\pi) e^{-i\omega t} \hat f(\omega)$. Equation \eqref{eq:eom_I} transforms into
\begin{equation}
  -m\omega^2\hat{\bb{u}}_I = - \bb{K}_I \hat{\bb{u}}_I - \sum_{J\neq I} \bb{V}_{IJ} \hat{\bb{u}}_J + i m \gamma_I \omega \hat{\bb{u}}_I + \hat{\xi}_I. \label{eq:eom_f}
\end{equation}
Rearrangement of Eq. \eqref{eq:eom_f} gives
\begin{equation}
 \hat{\bb{u}}_I(\omega) = \hatb{\bb{g}}_I(\omega) \left[\sum_{J\neq I} \bb{V}_{IJ}\hat{\bb{u}}_J(\omega) - \hat{\xi}_I(\omega)\right],
 \label{eq:ui}
\end{equation}
where the uncoupled Green's function is defined as
\begin{equation}
 \hatb{\bb{g}}_I (\omega) = \left[m \omega^2 - \bb{K}_I + im\gamma_I\omega\right]^{-1}.
\label{eq:leadgreen}
\end{equation}
The uncoupled Green's function includes damping self-energy $im\gamma_I\omega$ due to coupling to the heat baths.

Substituting Eq. \eqref{eq:ui} for $I={L,R}$ to \eqref{eq:eom_f} for $I=C$ gives
\begin{alignat}{2}
 -&m\omega^2\hat{{\bb{u}}}_C(\omega)  \notag \\ 
  &= -\bb{K}_C \hat{\bb{u}}_C (\omega) -\sum_{I=L,R} \bb{V}_{CI} \hatb{\bb{g}}_I(\omega) \notag \\
  &\quad \times \left[\bb{V}_{IC}\hat{\bb{u}}_C(\omega) - \hat{\xi}_I(\omega)\right] + im \gamma_C \omega \hat{\bb{u}}_C (\omega) + \hat{\xi}_C(\omega)\\
 &= -\bb{K}_C \hat{\bb{u}}_C (\omega) - \sum_{I=L,R} [ \Sigma_I(\omega) \hat{\bb{u}}_C (\omega) - \hat{\eta}_I(\omega)] \notag \\
  &\quad + im\gamma_C \omega \hat{\bb{u}}_C (\omega) + \hat{\xi}_C(\omega).
 \label{eq:eom_om}
\end{alignat}
In the second line, we defined the lead self-energies
\begin{equation}
 \Sigma_I(\omega) = \bb{V}_{CI} \hatb{\bb{g}}_I(\omega) \bb{V}_{IC}
 \label{eq:sigmaI_def}
\end{equation}
and the lead-coupled Langevin noise terms
\begin{equation}
 \hat{\eta}_I(\omega) = \bb{V}_{CI} \hatb{\bb{g}}_I(\omega) \hat{\xi}_I(\omega).
 \label{eq:eta_def}
\end{equation}
The solution to Eq. \eqref{eq:eom_om} is 
\begin{equation}
 \hat{\bb{u}}_C(\omega) = -\hatb{\bb{G}}(\omega) \left[ \hat{\xi}_C(\omega)+\sum_{I=L,R} \hat{\eta}_I(\omega)  \right],
\label{eq:sol_uom}
\end{equation}
where the full Green's function of the center region is 
\begin{equation}
 \hatb{\bb{G}}(\omega) = \left[ m\omega^2 -\bb{K}_C  + im\gamma_C \omega-\sum_{I=L,R} \Sigma_I(\omega) \right]^{-1}.
\label{eq:GCC}
\end{equation}

Equations \eqref{eq:eta_def} and \eqref{eq:sol_uom} state that thermal fluctuations in the leads can propagate to the center region as described by the Green's function $\bb{g}_I(\omega)$ and coupling matrix $\bb{V}_{CI}$ and thereby introduce additional noise terms $\hat{\eta}_L$ and $\hat{\eta}_R$ in the center region. The self-energies $\Sigma_L$ and $\Sigma_R$ appearing in the Green's function \eqref{eq:GCC} describe the energy shift and broadening of the phonon energy levels in the center region due to the leaking of phonons into the leads. The self-energies of the semi-infinite leads can be determined by using, e.g., the recursive decimation routine by Lopez and Sancho \cite{lopezsancho85}. 

By integrating out the leads, we have effectively replaced the lead coordinates by the noise terms $\eta_I$ and the accompanying self-energies $\Sigma_I$. In the next section, we derive the fluctuation-dissipation relation connecting the statistical properties of $\hat{\eta}_I$ to $\textrm{Im}[\Sigma_I(\omega)]$.

\subsection{Noise power spectra}
\label{sec:noise_properties}
 The Langevin noise operators $\xi_I$ appearing in Eq. \eqref{eq:eom_I} act as stochastic sources of thermal fluctuations due to coupling to the local heat bath \cite{weiss,ford88,dhar06}. They are Gaussian random variables with zero mean and covariance related to bath temperature. To calculate the statistical averages of observables such as heat current in the center region, we need the covariances of both the bare Langevin noises $\hat{\xi}_C$ and the lead noise contributions $\hat{\eta}_I=\bb{V}_{CI} \hatb{\bb{g}}_I \hat{\xi}_I$. Since the bath temperatures in the center region depend on position $i$, we handle the noise terms originating from the center region and leads separately. In the following, we assume for notational convenience that each atom only has a single degree of freedom corresponding to displacement along, say, $x$-direction. In the general case, local bath at site $i$ will be coupled to displacements $(u_i^x,u_i^y,u_i^z)$ in different coordinate directions with a single temperature $T_i$, making the notation a bit more cumbersome but analogous.

The covariance of the noises produced by the local heat baths at sites $i$ and $j$ in the center region is, for $\hbar=k_B=1$, \cite{ford88,dhar06}
\begin{equation}
  \langle \hat{\xi}_{Ci}(\omega) \hat{\xi}_{Cj}(\omega') \rangle = 2\pi \delta(\omega+\omega') \Gamma_{ij}(\omega) \left[ f_B(\omega,T_i) +1 \right],
 \label{eq:xiCxiC}
\end{equation}
where the coupling function for Ohmic friction is $ \Gamma_{ij}(\omega)= 2m\gamma_C \omega \delta_{ij}$. This corresponds to the memoryless friction assumed in Eq. \eqref{eq:eom_I}, but more general couplings could be straightforwardly included as well. The friction parameter $\gamma_C$ determines the strength of coupling to the heat baths and can be interpreted as phonon decay rate \cite{li09jap}. The corresponding scattering time is $\tau_C=\gamma_C^{-1}$, which is independent of frequency for Ohmic baths. 

The term in braces, where the bath temperature $T_i$ appears in the Bose-Einstein function $f_B(\omega,T_i)=[\exp(\omega/T_i)-1]^{-1}$, can be written in the more transparent form $f_B(\omega,T_i)+1/2+1/2$, where the first term is the thermal phonon occupation number, the second term comes from zero-point fluctuations and the last term reflects the non-commutative quantum nature of the Langevin operators \cite{wang07}. One can write the term as the sum of odd and even functions in $\omega$ as $f_B(\omega,T_i)+1=\coth(\omega/2T_i)/2+1/2$, and it turns out that the additional factor of $1/2$ cancels in all integrals over frequency after proper symmetrization.

For the noise operators in the leads, the temperatures of the baths have prescribed values $T_L$ and $T_R$, which do not depend on position. Therefore, we can write the covariance directly in matrix form as ($I\in\{L,R\}$)
\begin{equation}
  \langle \hat{\xi}_{I}(\omega) \hat{\xi}_{I}(\omega')^T \rangle = 2\pi \delta(\omega+\omega')  \tilde{\Gamma}^I (\omega) \left[ f_B(\omega,T_I)+1\right],
 \label{eq:xiIxiI}
\end{equation}
which is useful in the following calculations. Here, the coupling function matrix $\tilde{\Gamma}^I$ is a diagonal matrix with elements $\tilde{\Gamma}^I_{ij}(\omega) = 2m\gamma_I \omega \delta_{ij}$. 

Using $\langle \hat{\xi}_I(\omega)\rangle)=0$ and Eqs. \eqref{eq:eta_def} and \eqref{eq:xiIxiI}, we see that the noise terms $\hat{\eta}_I$ originating from the leads satisfy $\langle \hat{\eta}_I (\omega)\rangle=0$ and
\begin{equation}
 \langle \hat{\eta}_I(\omega) \hat{\eta}_I(\omega')^T \rangle = 2\pi \delta(\omega+\omega') \bb{I}_I(\omega),
\label{eq:eta_I}
\end{equation}
with the power spectrum
\begin{equation}
 \bb{I}_I(\omega) =  \bb{V}_{CI} \hatb{\bb{g}}_I(\omega)\tilde \Gamma^I(\omega)  \hatb{\bb{g}}_I(-\omega) \bb{V}_{IC} \left[f_B(\omega,T_I)+1\right], \label{eq:Ii1}
\end{equation}
where we noted that the Green's function $\bb{g}_I(\omega)$ is symmetric, since the spring matrix $\bb{K}_I$ is symmetric. A straightforward calculation shows that
\begin{alignat}{2}
 \hatb{\bb{g}}_I(\omega) \tilde{\Gamma}^I(\omega) \hatb{\bb{g}}_I(-\omega) = i[ \hatb{\bb{g}}_I(\omega) - \hatb{\bb{g}}_I(\omega)^*],
 \label{eq:gminusg}
\end{alignat}
so Eq. \eqref{eq:Ii1} becomes
\begin{alignat}{2}
 \bb{I}_I(\omega) &= i \bb{V}_{CI} \left(\hatb{\bb{g}}_I(\omega) - \hatb{\bb{g}}_I(\omega)^*\right)  \bb{V}_{IC} \left[f_B(\omega,T_I)+1\right]\\
  &= - 2 \bb{V}_{CI} \textrm{Im}\left[\hatb{\bb{g}}_I(\omega)\right] \bb{V}_{IC}  \left[f_B(\omega,T_I)+1\right] \\
  &= - 2 \textrm{Im}[\Sigma_I(\omega)] \left[f_B(\omega,T_I)+1\right],
\end{alignat}
where we used the definition \eqref{eq:sigmaI_def}. Defining the lead coupling function
\begin{equation}
 \Gamma^I(\omega) = -2 \textrm{Im}[\Sigma_I(\omega)],
\label{eq:gammaIdef}
\end{equation}
we see that the power spectrum of the noise caused by the leads can be written as
\begin{equation}
\langle \hat{\eta}_I(\omega) \hat{\eta}_I(\omega')^T \rangle = 2\pi \delta(\omega+\omega') \Gamma^I(\omega) \left[f_B(\omega,T_I)+1\right]. \label{eq:Iom}
\end{equation}
Equation \eqref{eq:Iom} is analogous to Eqs. \eqref{eq:xiCxiC} and \eqref{eq:xiIxiI} except for the form of the coupling matrix $\Gamma^I(\omega)$, now defined using the self-energy of the lead as shown in Eq. \eqref{eq:gammaIdef}. Equation \eqref{eq:Iom} is one of the main results of this paper, showing that an atomic reservoir (lead) coupled to local heat baths at prescribed temperature can be represented by noise and dissipation terms related by a fluctuation-dissipation relation. In contrast to previous works \cite{dhar03,dhar06}, our model assumes from the beginning that there is damping everywhere in the system. This results, e.g., in a lead Green's function \eqref{eq:leadgreen} that includes an additional self-energy term $im\gamma\omega$, in contrast to the ballistic lead Green's function defined, e.g., below Eq. (2.5) in \cite{dhar06}. Simply adding the damping to the Green's function used in Ref. \cite{dhar06} would not give a consistent mathematical picture of the situation, since the presence of damping also introduces thermal noise through the fluctuation-dissipation relation and thereby modifies the equations of motion.

Solution \eqref{eq:sol_uom} combined with the noise correlations \eqref{eq:xiCxiC} and \eqref{eq:Iom} allows us to calculate the thermal averages of all observables of interest.

\subsection{Heat flow to baths}

\label{sec:heatflowtobaths}
In the Heisenberg-Langevin equation of motion \eqref{eq:eom_om}, the friction and stochastic force terms induce energy exchange with the heat bath. The energy exchange rate can be calculated from the time derivative of local energy \cite{hardy63}. A natural definition for the local Hamiltonian of atom $i$ in the center region is
\begin{equation}
 h_i = \frac{p_i^2}{2m} + \frac{1}{2} \sum_{j} u_i K_{ij}u_j.
 \label{eq:hi_def}
\end{equation}
In Eq. \eqref{eq:hi_def} and from now on, we drop the index $C$ describing the center region, since the lead coordinates do not appear anymore. Using the equation of motion \eqref{eq:eom_I}, the symmetrized time derivative taking into account the non-commutativity of $u_i$ and $p_i$ can be calculated to be
\begin{alignat}{2}
 \dot {h}_i &= \frac{1}{2} \left\{\dot{p}_i,\frac{p_i}{m} \right\} + \frac{1}{4} \sum_j K_{ij} \left( \left\{\dot{u}_i,u_j \right\} +  \left\{u_i,\dot{u}_j \right\} \right) \label{eq:dothi1} \\
  &= - \frac{1}{4} \sum_{j}K_{ij} \left(\left\{\dot{u}_i,u_j\right\}  - \{u_i,\dot{u}_j\} \right) \notag \\
  &\quad - \left( m\gamma\dot{u}_i^2 - \frac{1}{2}\{\dot{u}_i,\xi_i\} \right). \label{eq:dothi}
\end{alignat}
Here $\{A,B\}=AB+BA$ is the anti-commutator and for simplicity, we have assumed that the particle is not at the boundary so that it is not directly coupled to the leads. The term inside the sum in Eq. \eqref{eq:dothi} is the heat current flowing from site $i$ to site $j$ and the second term in parentheses is the heat current
\begin{equation}
 Q_i =  m\gamma\dot{u}_i^2 - \frac{1}{2} \left[\dot{u}_i\xi_i+\xi_i\dot{u}_i\right]
\label{eq:Qi_def}
\end{equation}
flowing to the local heat bath at site $i$. 

As shown in the appendix, the statistical average of the heat current is formed as a sum over the contributions of the left ($J=L$) and right ($J=R$) leads and each local heat bath ($J\in\{1,2,\dots,N_C\}$) as
\begin{alignat}{2}
 \langle Q_i \rangle &= \int_0^{\infty} \frac{d\omega}{2\pi} 2 \gamma m \omega^2 \sum_J \left[\bb{G}(\omega) \Gamma^J(\omega) \bb{G}(-\omega)^T \right]_{ii} \notag \\ &\quad \times [f_B(\omega,T_J)-f_B(\omega,T_i)] .
 \label{eq:qlandauer_1}
\end{alignat}
Here the sum over bath index $J\in\{L,R,1,2,\dots,N_C\}$ separately accounts for the contribution of each individually treated heat bath to the thermal balance at site $i$ as detailed in the appendix. The coupling matrix $\Gamma^J$ is defined by Eq. \eqref{eq:gammaIdef} for the lead heat baths ($J=L$ or $J=R$). For the local heat baths ($J\in\{1,2,\dots,N_C\}$), the only non-zero element of the coupling matrix $\Gamma^J$ is $\Gamma^J_{JJ}=2\gamma m\omega$. The term $[\mathbf{G}\Gamma^J\mathbf{G}^{\dagger}]_{ii}$ describes the thermal coupling between bath $J$ and site $i$. In the following, we refer to the bath at site $i$ simply as bath $i$.

Using the definition of $\Gamma^i$ for local heat baths, we can write Eq. \eqref{eq:qlandauer_1} for the heat flow to bath $i$ in the general form
\begin{equation}
 \langle Q_i \rangle = \int_0^{\infty} \frac{d\omega}{2\pi} \omega \sum_{J} \mathcal{T}_{iJ}(\omega) \left[f_B(\omega,T_J)-f_B(\omega,T_i) \right],
\label{eq:Q_landauer}
\end{equation}
where the transmission function between baths $i$ and $J$ is
\begin{equation}
  \mathcal{T}_{iJ}(\omega)= \textrm{Tr}\left[\Gamma^i(\omega) \bb{G}(\omega)\Gamma^{J} (\omega) \bb{G}(-\omega) \right].
 \label{eq:Tij}
\end{equation}
Equation \eqref{eq:Q_landauer} is also valid for the currents flowing to the leads, i.e. for the substitution $i\to L$ or $i\to R$, and the derivation proceeds analogously. In this case, one should use the equation of motion $\eqref{eq:eom_om}$ in the $\dot{p}_i$ term of \eqref{eq:dothi} to calculate the heat in-flow to center region by the noise term $\hat{\eta}_I$ and out-flow by the force term $\Sigma_I\hat{\bb{u}}(\omega)$. 

Equation \eqref{eq:Q_landauer} is the multiprobe Landauer-B\"uttiker formula \cite{buttiker92} for thermal transfer between several heat baths. Equation \eqref{eq:Tij} is the Caroli formula \cite{caroli71} for phonon transmission function, first derived by Mingo and Yang \cite{mingo03} from the mode picture and by Yamamoto and Watanabe \cite{yamamoto06} using Keldysh formalism. We have rederived the formula using local Langevin heat baths and thereby also included dissipative effects in the leads.

We point out that although the average heat current $\langle Q_i \rangle$ vanishes in the self-consistent temperature configuration for all local heat baths in the center region, the spectral heat current
\begin{alignat}{2}
 \langle \hat{Q}_i(\omega) \rangle &= 2 \gamma m  \omega^2 \sum_J \left[\bb{G}(\omega) \Gamma^J(\omega) \bb{G}(-\omega)^T \right]_{ii} \notag \\ &\quad \times [f_B(\omega,T_J)-f_B(\omega,T_i)]
\end{alignat}
to a local heat bath is generally non-zero. For example, a bath may have a net in-flow of high-energy phonons, but then there must be a corresponding net out-flow of low-energy phonons. These non-zero spectral currents lead to the redistribution of phonon energies inside the structure, similarly to the full non-equilibrium Green's function formalism where generally $\langle Q_L(\omega) \rangle \neq - \langle Q_R(\omega) \rangle$ \cite{mingo06}.

In the limit of \bbf{vanishing couplings to local heat baths}, $\gamma \to 0^+$, $\gamma_I\to 0^+$, the lead and center region Green's functions \eqref{eq:leadgreen} and \eqref{eq:GCC} region reduce to their ballistic counterparts and the only non-zero transmission function is $\mathcal{T}_{LR}(\omega)$. Equation \eqref{eq:Q_landauer} reduces to
\begin{equation}
 \langle Q_R \rangle = \int_0^{\infty} \frac{d\omega}{2\pi} \omega  \mathcal{T}_{LR}(\omega) \left[f_B(\omega,T_L)-f_B(\omega,T_R) \right]
\end{equation}
for the current flowing to the right reservoir. This is the two-probe Landauer-B\"uttiker formula for ballistic phonon transfer, derived earlier by various methods \cite{rego98,angelescu98, mingo03,segal03,dhar03,yamamoto06}.

\subsection{Physical interpretation of the bath temperatures}
\label{sec:temperature_properties}
For the classical self-consistent thermal bath models, the requirement of zero net energy exchange with the local baths was enforced by requiring the bath temperatures to be equal to the local kinetic temperatures \cite{bolsterli70}.  The present model allows finding a fully quantum interpretation for the self-consistent bath temperatures. To this end, we first note that the heat current flowing to a self-consistent reservoir at site $i$ can be written in the form 
\begin{equation}
\langle Q_i \rangle =\gamma \left\{2 \langle e_i^{kin} \rangle - \int_0^{\infty} \frac{d\omega}{2\pi}\omega D_i(\omega) \left[ f_B(\omega,T_i)+\frac{1}{2}\right] \right\},
 \label{eq:QIekin}
\end{equation}
where the local kinetic energy is $e_i^{kin} = m \dot{u}_i^2/2$ and the local density of states (LDOS) is defined as $D_i(\omega) = -4\omega m \textrm{Im}[G_{ii}(\omega)]$ \cite{zhang07}. This form results from noting that the first term of Eq. \eqref{eq:Qi_def} is $\gamma m  \dot{u}_i^2 = 2\gamma e_i^{kin}$, and the second term follows from Eq. \eqref{eq:a10} by using the definition of LDOS and dropping the odd term that cancels out in the integration. The self-consistency criterion $\langle Q_i \rangle = 0$ then reduces to the requirement
\begin{equation}
 2\langle e_i^{kin}\rangle = \int_0^{\infty} \frac{d\omega}{2\pi}\omega D_i(\omega) \left[ f_B(\omega,T_i)+\frac{1}{2}\right].
\label{eq:ekin_dfb}
\end{equation}
The left-hand side of Eq. \eqref{eq:ekin_dfb} can be interpreted as the total energy at site $i$ consisting of the kinetic and elastic energies, which are equal in a statistical-mechanical system according to the virial theorem \cite{goldstein}. Virial theorem is, of course, rigorously valid only for the total kinetic and interaction energies at thermal equilibrium. The right-hand side is the total vibrational energy of an oscillator at temperature $T_i$. Equation \eqref{eq:ekin_dfb} gives a very natural interpretation to the self-consistent bath temperature $T_i$ as a measure of energy located at site $i$. 

Note that for Ohmic baths, the integrals in Eqs. \eqref{eq:QIekin} and \eqref{eq:ekin_dfb} actually diverge, because the density of states scales as $D(\omega)\sim -\omega\textrm{Im}[1/(\omega^2+i\gamma m\omega)] \sim \omega^{-2}$ for $\omega \to \infty$, resulting in a logarithmic divergence in the zero-point term. The divergence is, however, cancelled by an identical term in $\langle e_i^{kin} \rangle$, making $T_i$ well-defined. 

In the classical limit, Eq. \eqref{eq:ekin_dfb} reduces to 
\begin{equation}
 \langle e_i^{kin}\rangle = \frac{T_i}{2} \int_0^{\infty} \frac{d\omega}{2\pi}D_i(\omega) = \frac{1}{2} T_i,
\label{eq:ekin_classical}
\end{equation}
where we used the sum rule $\int_0^{\infty} (d\omega/2\pi)D_i(\omega)=1$. This sum rule has been proven for the electronic case \cite{mahan} and the proof for phonons is analogous. Equation \eqref{eq:ekin_classical} can be interpreted as the local equipartition theorem analogous to the statistical mechanical equipartition theorem $\langle e_{kin} \rangle=N_fT/2$, where $N_f$ is the number of degrees of freedom in the system. Relation \eqref{eq:ekin_classical} is routinely used as the definition of local temperature in classical molecular dynamics simulations \cite{lepri03}. Equation \eqref{eq:ekin_dfb} suggests a similar definition for quantum systems.

\subsection{Solution of the Boltzmann transport equation in 1D chain}
\label{sec:bte_theory}
In a sense, self-consistent thermal bath model (SCTB) can be thought of as the fully wave enabled extension of the gray approximation \cite{majumdar93,murthy05} to the Boltzmann transport equation (BTE). Therefore it is instructive to compare the results obtained from BTE and SCTB under conditions where the wave-effects are negligible and there are no reflections between the chain and the reservoirs. For the simple one-dimensional string of length $L$, BTE in the continuum approximation reads
\begin{subequations}
 \begin{align}
 v(\omega) \frac{\partial n_+(x,\omega)}{\partial x} &= - \frac{n_+(x,\omega)-n_0(x,\omega)}{\tau} \label{eq:bte1} \\ 
 - v(\omega)\frac{\partial n_-(x,\omega)}{\partial x} &= - \frac{n_-(x,\omega)-n_0(x,\omega)}{\tau} \label{eq:bte2},
 \end{align}
\end{subequations}
where $n_+(x,\omega)$ and $n_-(x,\omega)$ are the distribution functions for states with positive and negative group velocities, respectively. The thermal boundary conditions are $n_+(0,\omega)=f_B(\omega,T_L)$ and $n_-(L,\omega)=f_B(\omega,T_R)$. Distribution functions relax towards the average distribution $n_0=(n_++n_-)/2$ with relaxation time $\tau$. Mode dispersion in a one-dimensional chain is $\omega(q)=2\omega_0\sin(qa/2)$, where $a$ is the lattice constant. Note that the dispersion of the discrete chain is used in Eqs. \eqref{eq:bte1} and \eqref{eq:bte2} as usual (see, e.g., Ref. \cite{minnich11}). Mode velocity is $v(\omega) \equiv d\omega/dq =  a\omega_0 \sqrt{1-(\omega/2\omega_0)^2}$ and the density of states is $D(\omega)\equiv dq/d\omega=v(\omega)^{-1}$. 

The solution of BTE is
\begin{subequations}
\begin{align}
 n_+(x,\omega) &= f(\omega,T_L) - C(\omega) \frac{x}{2\Lambda(\omega)} \\
 n_-(x,\omega) &= f(\omega,T_L) - C(\omega) \left[1 + \frac{x}{2\Lambda(\omega)} \right],
\end{align}
\end{subequations}
where $\Lambda(\omega) = \tau v(\omega)$ is the scattering length and $C(\omega)= [f_B(\omega,T_L)-f_B(\omega,T_R)]/[1+L/2\Lambda(\omega)]$. The solution results in the heat current (again for $\hbar=k_B=1$)
\begin{alignat}{2}
  Q &= \int_0^{2\omega_0} \frac{d\omega}{2\pi} \omega \underbrace{v(\omega)D(\omega)}_{1} [ n_+(x,\omega)-n_-(x,\omega)] \\
    &= \int_0^{2\omega_0} \frac{d\omega}{2\pi} \omega \frac{1}{1+L/2\Lambda(\omega)} [f_B(\omega,T_L)-f_B(\omega,T_R)],
 \label{eq:jeff}
\end{alignat}
which can be interpreted as Landauer-B\"uttiker current with the effective transmission function 
\begin{equation}
 T_{eff}(\omega)  = \frac{1}{1+L/2\Lambda(\omega)}.
 \label{eq:Teff}
\end{equation}
For $L\gg \Lambda(\omega)$ and classical statistics ($T\gg\omega_0$), the thermal conductivity $\kappa=\lim_{\Delta T \to 0} \frac{QL}{\Delta T}$ derived from Eq. \eqref{eq:jeff} coincides with the expression obtained for the classical self-consistent heat bath model \cite{dhar06,li09jap} when the BTE relaxation time $\tau$ is identified with the inverse of the bath coupling constant $\gamma$. This shows the similarity of SCTB and BTE models in infinite classical systems.

\section{Solving the self-consistent equations}
\label{sec:methods}
The bath temperatures in the center region are determined by demanding that the average heat current $\langle Q_i \rangle$ to the local heat baths $i\in\{1,2,\dots,N_C\}$ given by Eq. \eqref{eq:qlandauer_1} vanishes. Since the bath temperatures appear in the Bose-Einstein functions, the equations are non-linear and the temperatures must be solved by using iterative methods or by resorting to linearizing approximations. We use both approaches and compare solutions obtained from the full non-linear equations with linear and classical approximations. Solutions of the non-linear equations are calculated using the Newton-Raphson method and in some cases an integration method based on the existence of a steady state towards which the system evolves.

The Newton-Raphson method has previously been used to solve the SCTB equations for 1D chains \cite{bandyopadhyay11} and is quite an efficient and reliable method for solving more general problems as well, especially when the linearized solution is used as the initial guess. However, each iteration of the Newton-Raphson method requires evaluating $N_C^2$ frequency integrals, which makes the method heavy for large systems. 

A slightly different and potentially better-scaling method for solving the equations can be found by writing a set of equations for the time evolution of the temperatures of the local baths and letting the system evolve towards the steady state. If the reservoir is imagined to have heat capacity $C_i$, the temperature $T_i$ of the reservoir changes due to the in-flow or out-flow of thermal current and obeys the differential equation
\begin{equation}
 \frac{dT_i(t)}{dt} = \frac{1}{C_i}\langle Q_i(T_1,\dots,T_{N_C}) \rangle ,
 \label{eq:dTdt}
\end{equation}
where the time $t$ is now macroscopic time such that any fluctuations in $Q_i$ vanish in the timescale of interest. This is an ordinary differential equation (ODE) of first order that evolves toward a steady-state where the bath temperatures satisfy the self-consistent temperature condition $\langle Q_i( T_1, \dots,T_{N_C})\rangle = 0 \quad \forall i\in\{1,2,\dots,N_C\}$. We call the method of integrating Eq. \eqref{eq:dTdt} the ODE method. In addition to having an intuitive physical interpretation as transient time evolution of the heat bath temperatures, ODE method has the advantage that at each time step, one only needs to calculate $N_C$ frequency integrals to calculate the time-derivative $(dT_1/dt,\dots,dT_{N_C}/dt)$. For large systems, the method could therefore provide a good alternative to the Newton-Raphson method. The heat capacity $C_i$ simply affects the time-scaling in Eq. \eqref{eq:dTdt} and can be included in the time variable $t$. 

The full solution of the nonlinear equations can be avoided by two common approximations that provide a linear set of equations for the bath temperatures \cite{segal09}. Linear response approximation is based on the assumption that temperature differences are small, allowing one to make in Eq. \eqref{eq:qlandauer_1} the substitution
\begin{equation}
 f_B(\omega,T_J) - f_B(\omega,T_i) \to \frac{\partial f_B}{\partial T}(\omega,T_m)(T_J-T_i),
\end{equation}
where the mean temperature is $T_m=(T_L+T_R)/2$. The linearization typically produces too low bath temperatures compared to the exact results \cite{bandyopadhyay11}. By considering a single-site model, we have traced this feature back to the fact that the second derivative of the Bose-Einstein function with respect to temperature is strictly positive.

Unlike the linear-response approximation, the classical approximation
\begin{equation}
 f_B(\omega,T_J)-f_B(\omega,T_i) \to \frac{1}{\omega} (T_J-T_i).
\end{equation}
makes the self-consistent equations linear also in lead temperatures, so the scaling of lead temperatures by a constant simply scales the self-consistent bath temperatures by the same factor. 

Both linearizations exclude any non-linear effects such as thermal rectification \cite{segal09} and produce more symmetric temperature profiles than the non-linear equations due to the equivalence of mapping $T_i \to T_L+T_R-T_i$ and the spatial reflection of the structure. 

\section{Numerical results}

\begin{figure}
 \includegraphics[width=8.6cm]{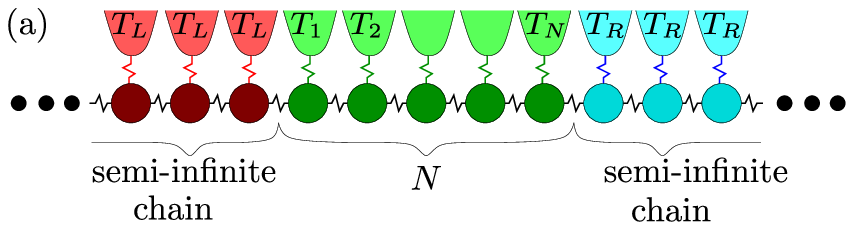}
 \includegraphics[width=8.6cm]{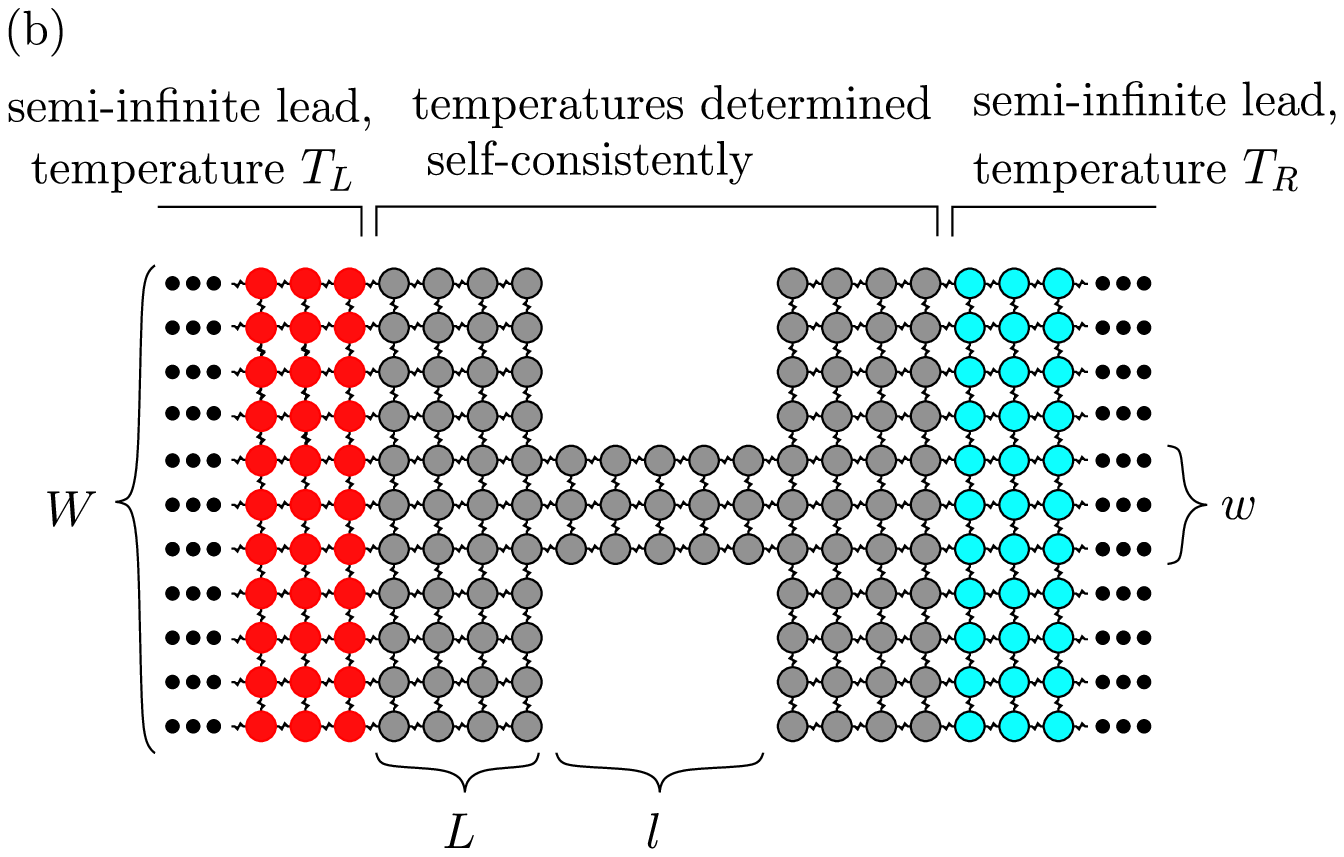}
 \caption{(Color online) Illustration of the systems studied in Secs. \ref{sec:rg_chain} and \ref{sec:2d_junction}: (a) a chain of length $N$ connected to two semi-infinite chains, (b) a constriction of width $w$ and length $l$ between two leads of width $W$. $L$ layers of atoms in the leads are included in the self-consistent calculation to account for the gradual temperature drop near the constriction.}
 \label{fig:chain_rg}
\end{figure}

To highlight the pertinent physics and the properties of the exact and approximate solutions of the self-consistent equations, we study in more detail the thermal conduction and temperature profiles in two structures shown in Fig. \ref{fig:chain_rg}. In the one-dimensional setup of Fig. \ref{fig:chain_rg}(a),  the temperatures are determined self-consistently in the center region consisting of a chain of $N$ atoms. The chain is connected to two semi-infinite chains interacting with heat baths at constant temperatures $T_L$ and $T_R$ so that there is no geometric scattering and phonon flow is reduced only by interactions with the local heat baths. The setup reduces to the Rubin-Greer geometry \cite{rubin71} if the heat baths are removed.

In the two-dimensional constriction geometry of Fig. \ref{fig:chain_rg}(b), two wide leads are connected by a narrow constriction. The center region includes not only the constriction but also $L$ layers of lead atoms to account for the effects of temperature drop near the constriction. In both geometries, nearest neighbors are assumed to be connected by harmonic springs with spring constant $k=m\omega_0^2$, where $m$ is the mass of the atoms. Each atom has only a single degree of freedom corresponding to, e.g., the atomic displacement in the out-of-plane direction.

Unless otherwise stated, we set $\omega_0=1$ in the following so that dimensionless temperatures are in units of $\hbar \omega_0/k_B$ and thermal currents in units of $\hbar \omega_0^2$. The dimensionless friction parameter is then in units of $\omega_0$. The friction parameter in the leads is set equal to the friction in the central region, $\gamma_C=\gamma_L=\gamma_R=\gamma$. In Secs. \ref{sec:rg_chain}, \ref{sec:2d_junction} and \ref{sec:graphene}, all exact self-consistent temperature configurations are calculated using the Newton-Raphson iteration with the linear response temperatures used as the initial guess. Newton-Raphson and ODE methods are compared in Sec. \ref{sec:results_comparison}.  

\subsection{Rubin-Greer chain}
\label{sec:rg_chain}
\begin{figure}
 \includegraphics[width=8.6cm]{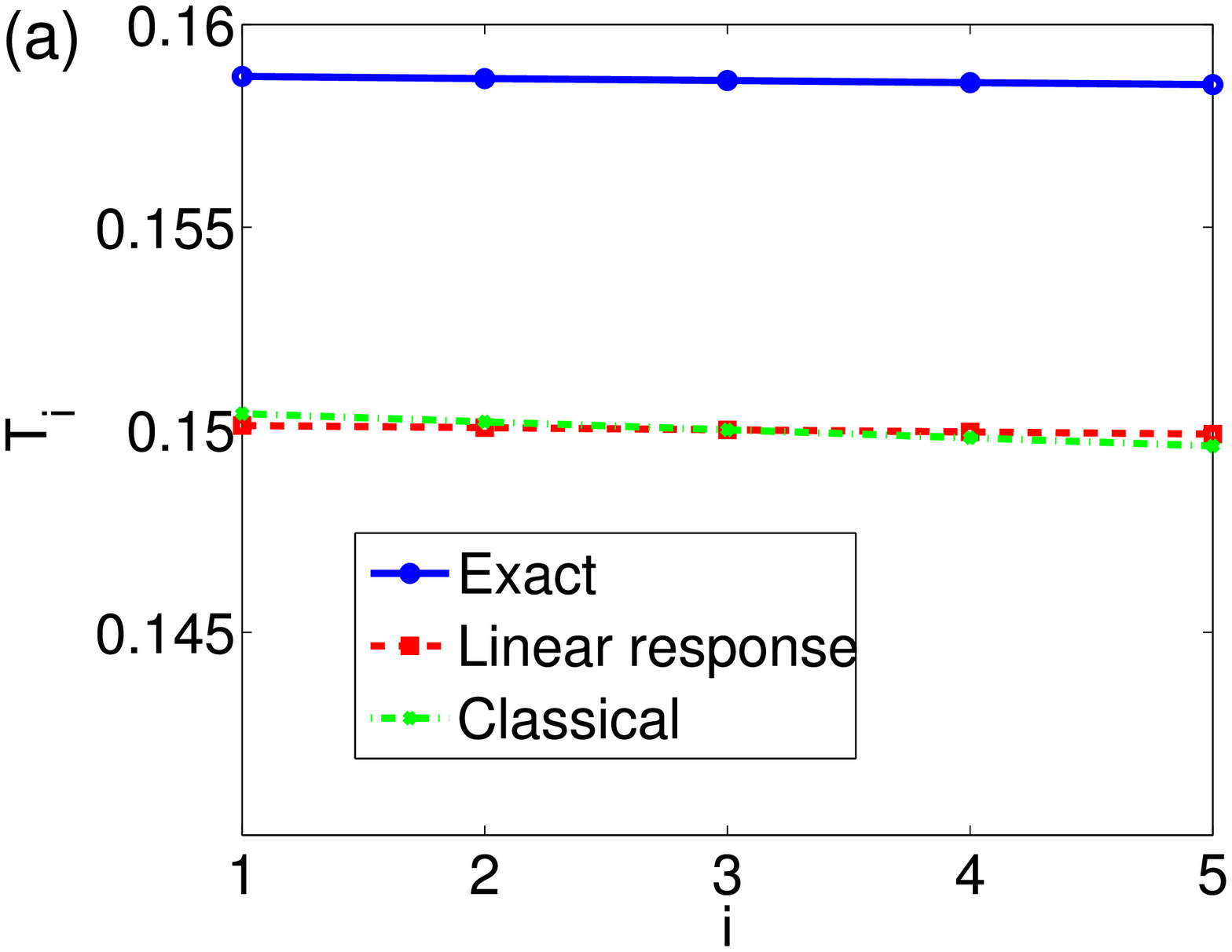}
 \includegraphics[width=8.6cm]{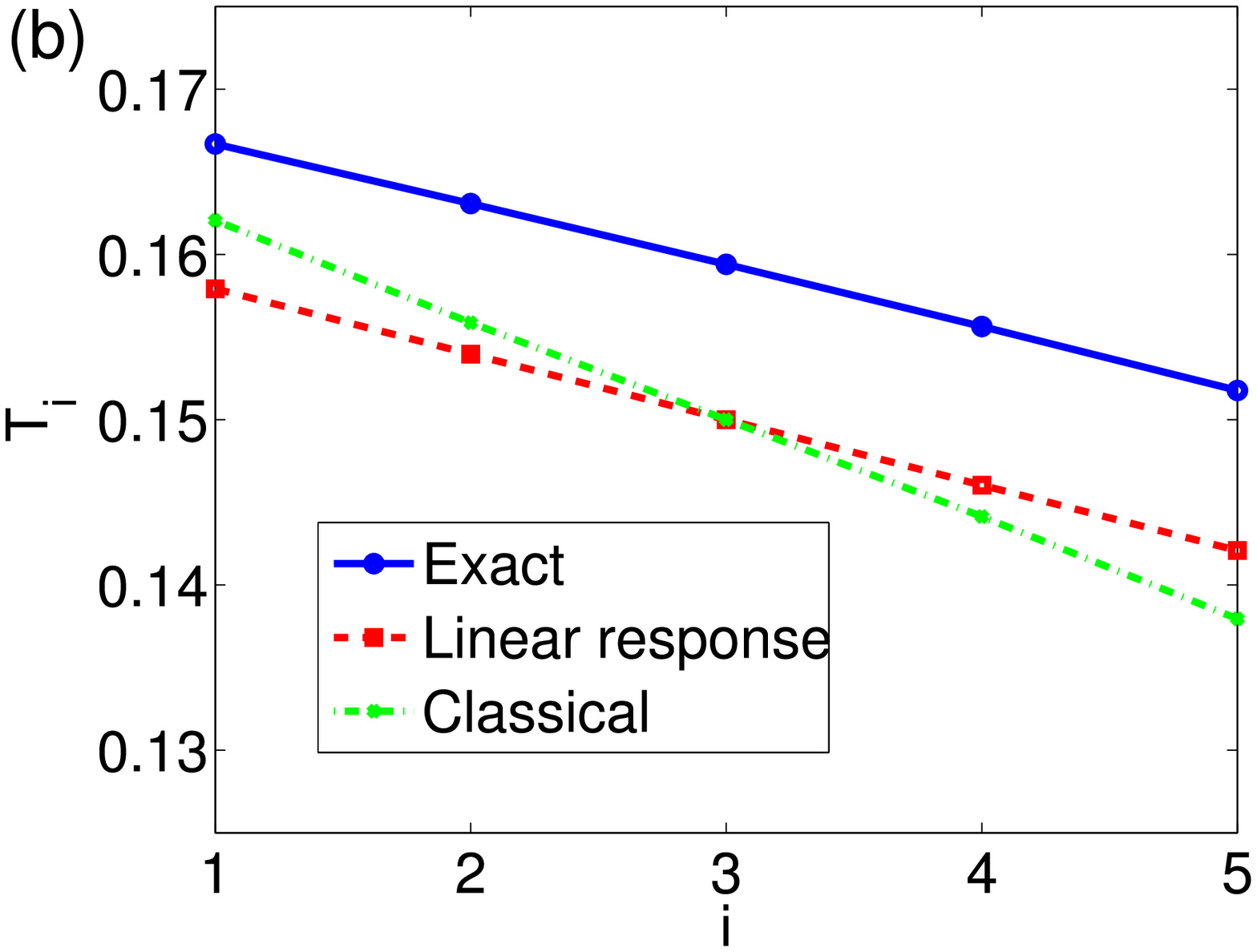}
 \caption{(Color online) Bath temperature profiles in a chain of length $N=5$ sandwiched between two semi-infinite leads at temperatures $T_L=0.2$ and $T_R=0.1$. Friction parameters are (a)  $\gamma=10^{-3}$, i.e. the system is nearly ballistic and (b) $\gamma=0.1$. Symmetry requires that $T_3=0.15$ for the linearized models, but the quantum exact temperature is higher due to the non-linearity of the Bose-Einstein function.}
 \label{fig:151212a_Ti}
\end{figure}

Due to the lack of geometric scattering in the Rubin-Greer setup of Fig. \ref{fig:chain_rg}(a), the setup serves as an ideal simplified model to compare the basic differences and similarities between the exact and approximate solutions of the self consistent problem. Figure \ref{fig:151212a_Ti} compares the self-consistent quantum exact, quantum linear and classical bath temperature profiles in a chain of length $N=5$ with friction parameters (a) $\gamma=10^{-3}$ and (b) $\gamma=0.1$. The lead temperatures are set to $T_L=0.2$, $T_R=0.1$. In the nearly ballistic system of Fig. \ref{fig:151212a_Ti}(a), all temperature profiles are nearly constants as a function of position, because coupling to baths is too weak for efficient thermalization. For increased damping in Fig. \ref{fig:151212a_Ti}(b), there is a clear temperature gradient due to interaction with the heat baths. The temperature gradients of the quantum exact and quantum linear response models are approximately the same, but the classical gradient is clearly larger. The most prominent feature in Figs. \ref{fig:151212a_Ti}(a) and \ref{fig:151212a_Ti}(b) is, however, that the quantum exact temperature is higher than the temperatures obtained in linear approximations, as noted also earlier for Ohmic leads \cite{bandyopadhyay11}.

\begin{figure}
 \includegraphics[width=8.6cm]{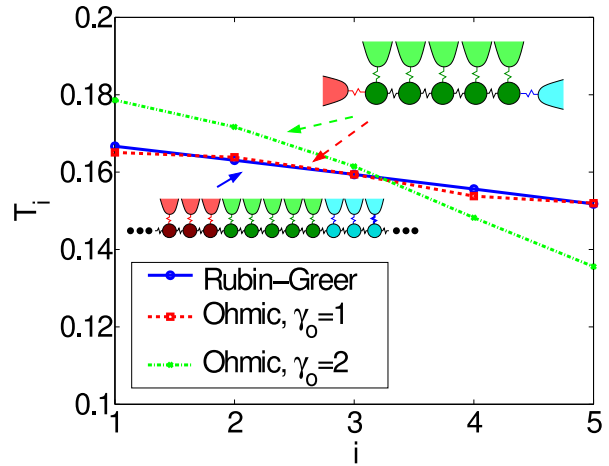}
 \caption{(Color online) Comparison of temperature profiles in a chain of length $N=5$ for two kinds of external reservoirs: Rubin-Greer leads [Fig. \ref{fig:chain_rg}(a)] and Ohmic reservoirs studied in Ref. \cite{bandyopadhyay11}. In contrast to Ref. \cite{bandyopadhyay11}, we also couple the particles at the ends of the chain to self-consistent baths. The reservoirs are at temperatures $T_L=0.2$ and $T_R=0.1$ and the coupling constant to self-consistent baths inside the chain is $\gamma=0.1$. In the Rubin-Greer setup, the coupling constant in the leads is $\gamma_L=\gamma_R=0.1$. In the case of Ohmic external reservoirs, coupling constant at the ends of the chain is denoted by $\gamma_o$. Choosing $\gamma_o=1$ reproduces the low-frequency self-energy of the Rubin-Greer chain, leading to similar temperature profiles at low temperatures. The geometries are shown in insets.}
 \label{fig:250513b}
\end{figure}
To highlight the difference of the Rubin-Greer setup to the Ohmic reservoirs studied in Ref. \cite{bandyopadhyay11}, we compare in Fig. \ref{fig:250513b} the temperature profiles for the two setups. In the low-frequency limit $\omega \to 0$, the self-energy of the semi-infinite Rubin-Greer chain is $\Sigma_{RG}(\omega)\approx -1-i\omega$ \cite{hopkins09}. The real part effectively means that the ends of the chain are free and not coupled to fixed particles as in Ref. \cite{bandyopadhyay11}. The imaginary part of the low frequency approximation of the Rubin-Greer chain self-energy can then be imitated by an Ohmic self-energy $\Sigma_o=-i\gamma_o\omega$ by choosing $\gamma_o=1$. Since only low-frequency phonons are excited at low temperature, the low frequency approximation is fairly accurate at low temperature and the temperature profiles are then expected to agree closely, as verified by Fig. \ref{fig:250513b} for end temperatures $T_L=0.2$, $T_R=0.1$. For $\gamma_o=2$, the temperature profile is steeper due to the stronger coupling to the external baths at the ends of the chain, which also introduces more dissipation in the system. At higher temperature, high-frequency phonons in the non-linear range of self-energy are excited as well, and the Ohmic coupling with $\gamma_o=1$ cannot reproduce the temperature profile as closely any more (not shown).

\begin{figure}
 \includegraphics[width=8.6cm]{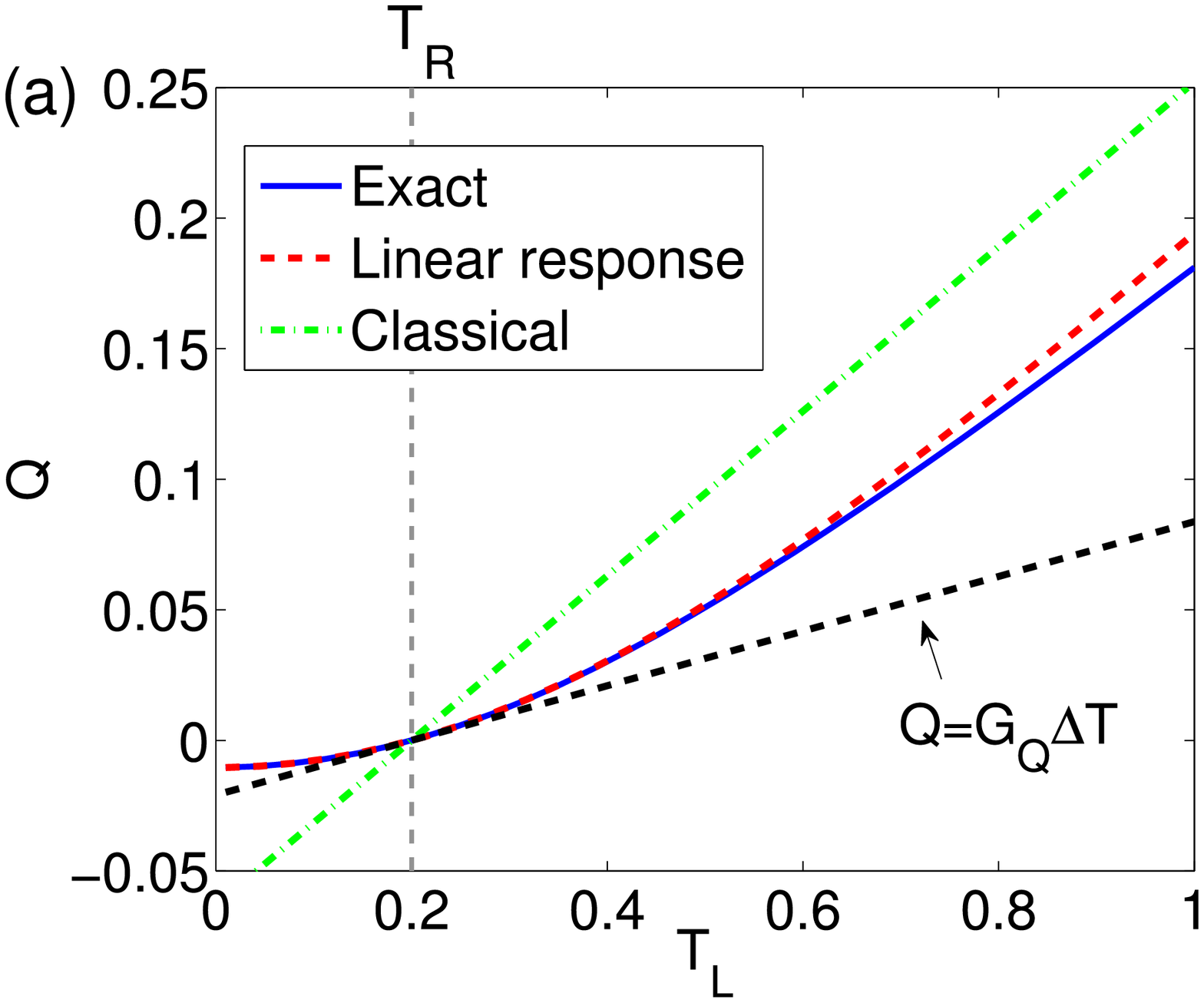}
 \includegraphics[width=8.6cm]{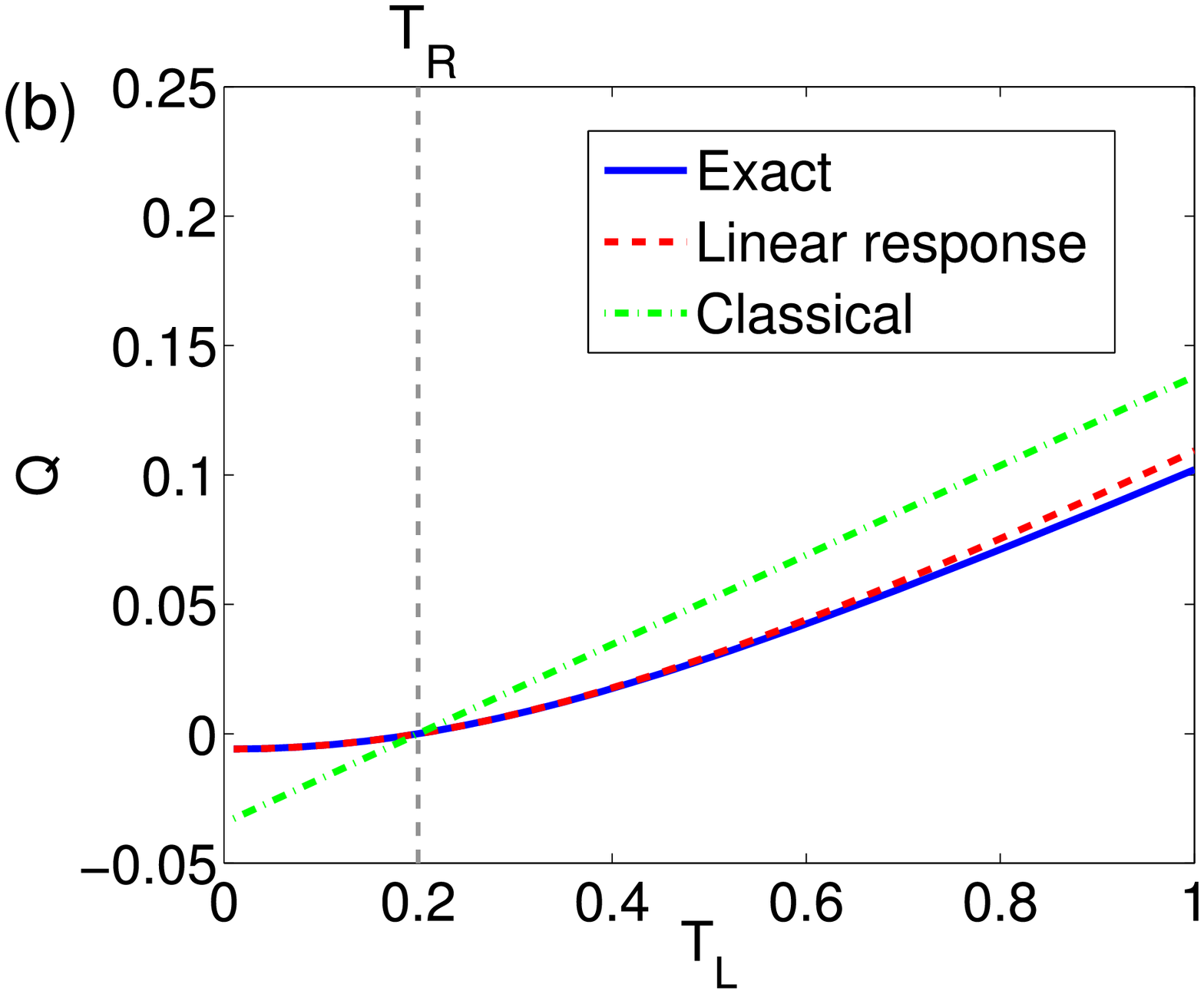}
 \caption{(Color online) Thermal current $Q$ as a function of lead temperature $T_L$ for fixed $T_R=0.2$. Chain length is $N=10$ and the friction parameters are (a) $\gamma=10^{-3}$ and (b) $\gamma=0.1$. With both weak and strong friction, the linear response approximation reproduces the exact current up to very high values of bias. In (a), current $Q=G_Q(T_L-T_R)$ corresponding to the quantum of thermal conductance $G_Q=\pi T/6$ with $T=0.2$ is also shown (black dashed).}
 \label{fig:chain_TL}
\end{figure}

Figure \ref{fig:chain_TL} shows the thermal current $Q\equiv \langle Q_R \rangle$ through the chain as a function of left lead temperature $T_L$ for fixed right lead temperature $T_R=0.2$. Friction parameters are set to (a) weak friction $\gamma=10^{-3}$ and (b) strong friction $\gamma=0.1$. The length of the self-consistently modeled chain is $N=10$. In the ballistic limit of Fig. \ref{fig:chain_TL}(a), the current flowing through the chain at low bias $T_L\approx T_R$ is equal to $Q=G_Q (T_L-T_R)$, where the quantum of thermal conductance \cite{rego98} is $G_Q=\pi k_B^2 T/6\hbar$, which reduces to $G_Q=\pi T/6$ in present units. When friction is increased in Fig. \ref{fig:chain_TL}(b), currents decrease due to the phonon damping caused by the heat baths. With both weak and strong friction, the classical approximation strongly overestimates thermal current, but the linear response approximation is valid up to $T_L \lesssim 0.6$. The classical approximation also makes the current response fully linear.

\begin{figure}
 \includegraphics[width=8.6cm]{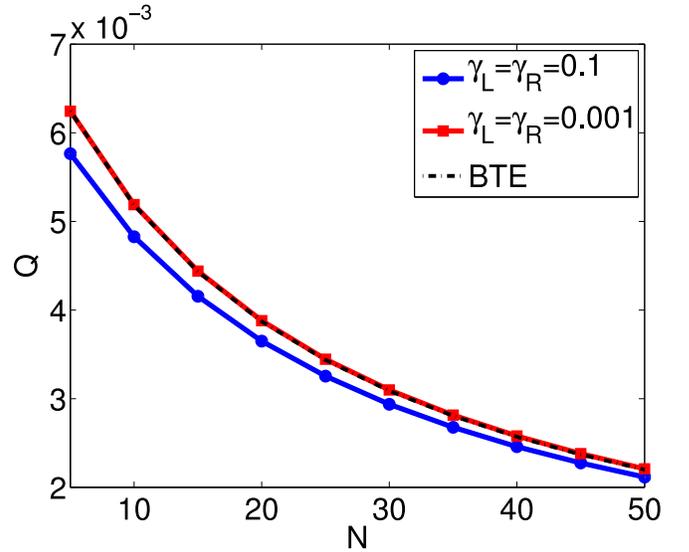}
 \caption{(Color online) Exact thermal current as a function of chain length $N$. The friction parameter in the center region is $\gamma=0.1$. The BTE current \eqref{eq:jeff} matches the self-consistent current if the leads are nearly ballistic, $\gamma_L=\gamma_R=0.001$. The bath temperatures in the left and right semi-infinite chains are $T_L=0.2$ and $T_R=0.1$, but the currents also agree at other temperatures for ballistic leads.}
 \label{fig:chain_length}
\end{figure}

Figure \ref{fig:chain_length} compares the thermal currents given by the self-consistent thermal bath (SCTB) solution and Boltzmann transport equation (BTE) solution \eqref{eq:jeff} as a function of chain length $N$. In the SCTB model, the friction parameter in the center region is $\gamma=0.1$ and in the leads $\gamma_L=\gamma_R=0.1$ or $\gamma_L=\gamma_R=0.001$. The string length $L$ in BTE is set to $L=(N+1)a$ to correspond to a chain of $N$ atoms and the relaxation time in the chain is $\tau=\gamma^{-1}=10$. The temperatures of the baths are set to $T_L=0.2$ and $T_R=0.1$ so that the system is in the non-linear low-temperature regime. As expected, the current decreases in both models as a function of chain length due to phonon decay in the chain. The BTE result matches the exact SCTB result perfectly, when the lead friction parameters $\gamma_L$ and $\gamma_R$ are small, i.e., the leads are assumed to be nearly ballistic and the center region friction parameter is tied by the relation $\gamma=\tau^{-1}$. The requirement of ballistic leads is natural, since we assumed that the phonon occupation in the leads is given by the Bose-Einstein distribution, which in SCTB model is exactly valid only in the limit of zero broadening, $\gamma_R=\gamma_L\to 0^+$. We have verified that the BTE and SCTB heat currents for small $\gamma_R=\gamma_L$ agree also at other temperatures. Figure \ref{fig:chain_length} also shows that increasing $\gamma_L$ and $\gamma_R$ in the leads, which increases scattering, slightly reduces the thermal current flowing through the center region. 

Despite the similarities between the predictions of BTE and SCTB for the simple 1D geometry, the models are not equivalent. For more complex geometries, the Green's function method, which contains full atomistic dynamics, wave effects and geometric scattering, is a drastic improvement over solving BTE under gray approximation.

The agreement of thermal currents between the SCTB model and BTE would have been very cumbersome to highlight, if the leads had been described by Ohmic reservoirs as in earlier works \cite{bandyopadhyay11} instead of Rubin-Greer chains. Because Ohmic baths at the ends of the chain would reflect some of the phonons back to the chain, the thermal boundary conditions for the distribution functions of right and left-moving phonons in the BTE formulation would have been different from simple Bose-Einstein functions.

\subsection{Constriction in two-dimensional lattice}
\label{sec:2d_junction}

In real systems, phonon transport is more complicated than in a one-dimensional chain due to, e.g., phonon reflections from boundaries. The new features arising from mode mismatch at contacts and other geometric factors will be studied in the constriction geometry of Fig. \ref{fig:chain_rg}(b), where the atoms are set in a square lattice such that a constriction of width $w$ and length $l$ connects two leads of width $W$. To account for the effects of temperature drop near the junction, $L$ atom layers in the leads closest to the constriction are also included in the self-consistent calculation. The constriction geometry has been studied earlier using molecular dynamics \cite{saha07, saaskilahti12}, but in contrast to molecular dynamics, the present methodology allows to include full quantum statistics in the phonon populations. From application point of view, constrictions are interesting due to their ability to act as thermal insulators, as noted in a recent experiment in GaAs point contacts \cite{bartsch12}. Although the present square lattice model is too primitive to accurately handle the experimental situation, our model could be used to gain insight into the local temperature profiles and diffusive effects inside the constriction.

\begin{figure}
 \includegraphics[width=8.6cm]{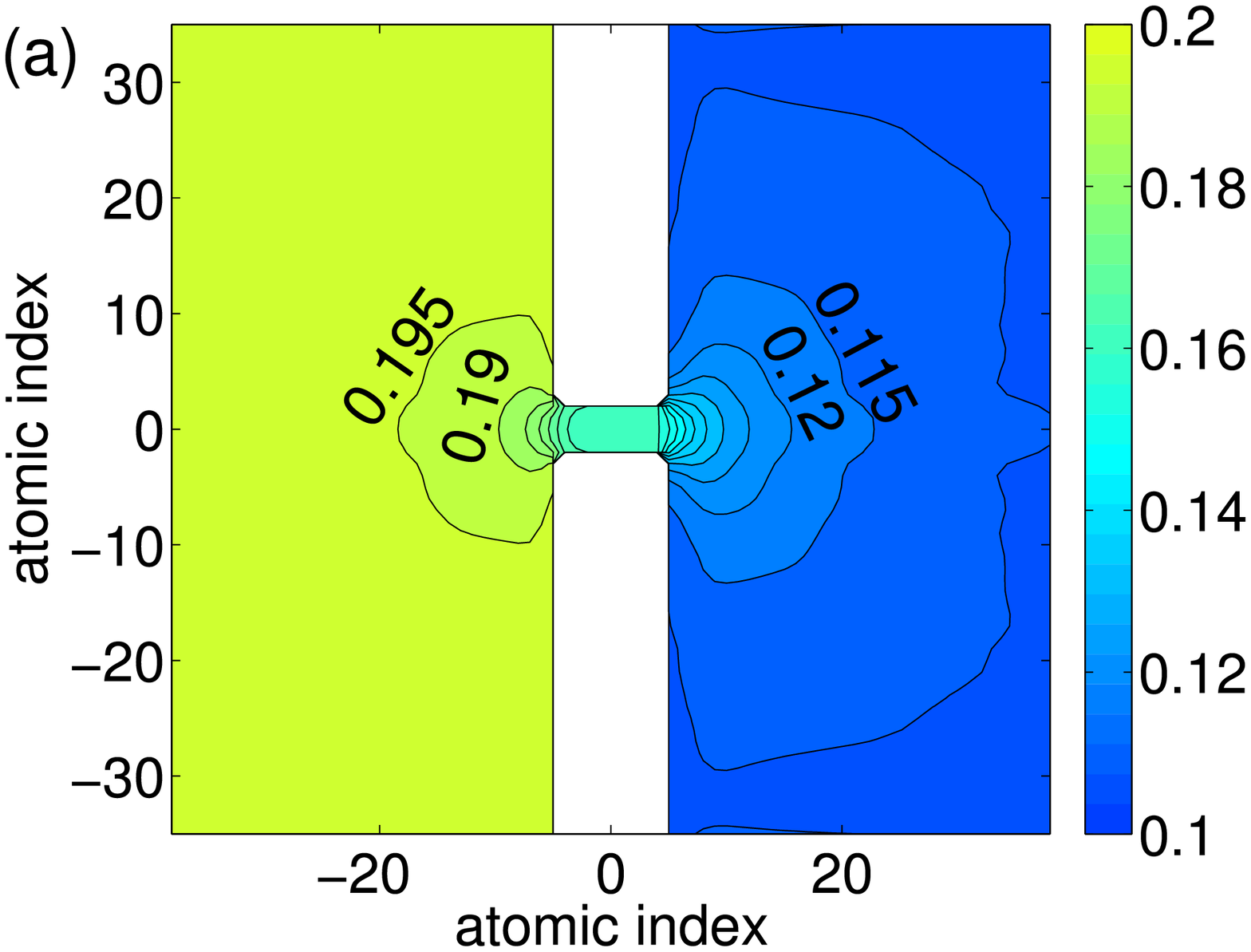}
 \includegraphics[width=8.6cm]{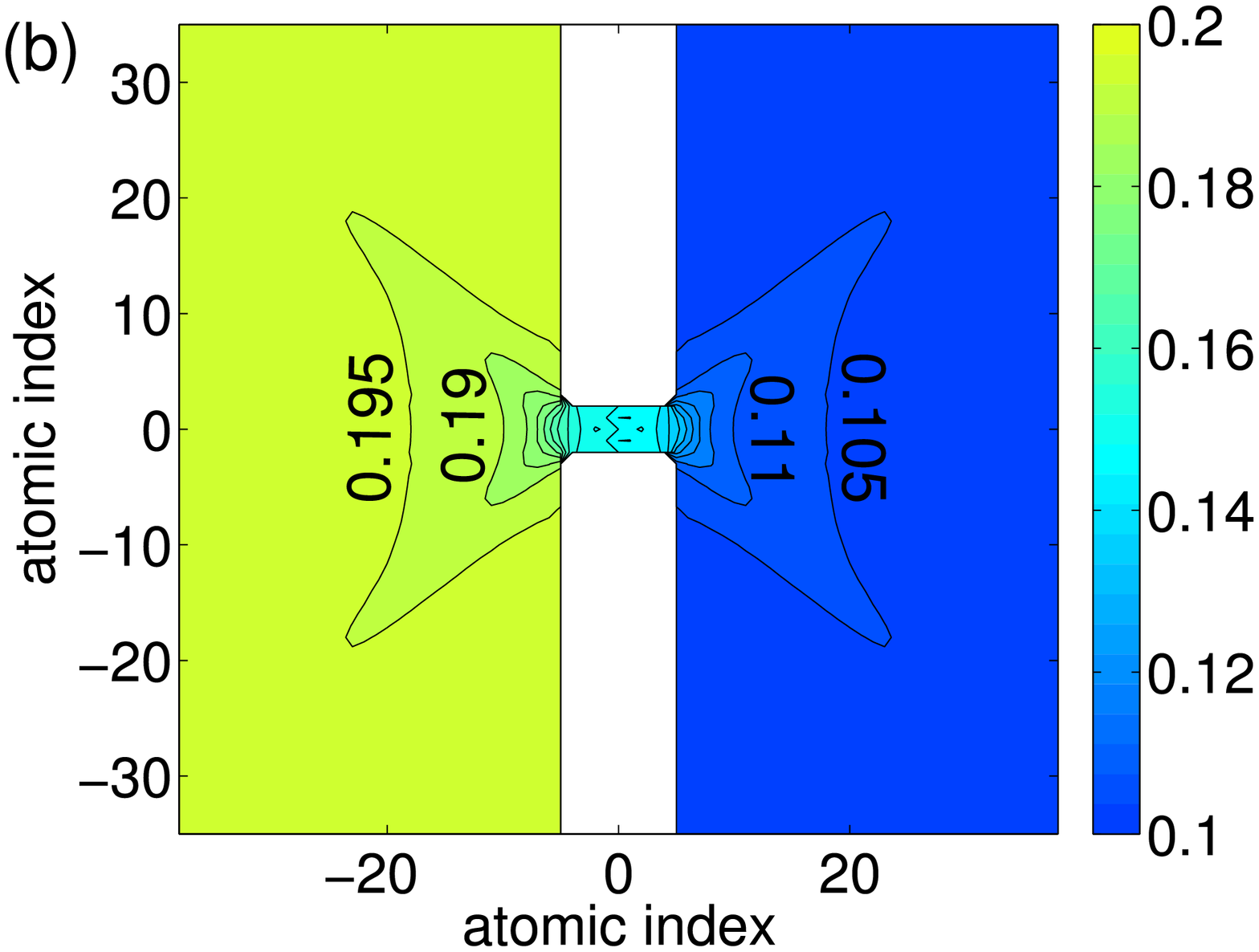}
 \caption{(Color online) Bath temperature profiles in a $w=5$, $l=9$ constriction coupled to leads of width $W=71$ and $L=35$ [see Fig. \ref{fig:chain_rg}(b)]. Lead temperatures are $T_L=0.2$ and $T_R=0.1$. Figures show (a) quantum exact and (b) classical self-consistent bath temperature profiles. Friction parameter is $\gamma=0.01$. The separation of isolines is $0.05$ and four contour lines are labeled for convenience.}
 \label{fig:180113a_Tis}
\end{figure}

Figure \ref{fig:180113a_Tis} shows the (a) quantum exact and (b) classical temperature profiles in a $w=5$, $l=9$ constriction coupled to leads of size $W=71$ and $L=35$ in the low-temperature ($T_L=0.2$, $T_R=0.1$) and nearly ballistic regime ($\gamma=0.01$). The asymmetry arising from the non-linearity of the self-consistency equations is very prominent in the quantum exact profile of Fig. \ref{fig:180113a_Tis}(a), as the temperature profile patterns in the left and right sides are visibly different. The junction temperature is approximately $0.17$, which is notably higher than the average temperature $0.15$. This is a similar effect as noticed in the previous section for the 1D chain: the mixing of the statistics of phonons at hot and cold temperatures results in a thermal population whose temperature is higher than the average temperature. 

In the classical case of Fig. \ref{fig:180113a_Tis}(b), on the other hand, symmetry of the self-consistent model requires that the temperature profile is symmetric with respect to spatial reflection and mapping $T_i\to T_L+T_R-T_i$. Therefore, the central part of the junction is at temperature $0.15$. In addition, the temperature profile in the bulk parts exhibits directional features at 45 degree angles with respect to the junction. These features have been observed also earlier for similar geometry in classical molecular dynamics simulations \cite{saaskilahti12}. In the quantum profile, these diagonal directional features are absent and the temperature profiles are more directed straight towards the leads. This feature is even more prominent for narrower constrictions (not shown). The difference between the quantum and classical profiles is most likely related to the transmission properties of high-energy phonons $(\omega \gtrsim T)$, whose populations are overestimated by classical statistics. Another major difference between quantum and classical statistics is that the currents flowing through the structure are $Q=84.1 \times 10^{-4}$ for quantum statistics and $Q=138 \times 10^{-3}$ for classical statistics, i.e., current is very strongly overestimated by the classical statistics in the low-temperature regime, as noted also for 1D chain [Fig. \ref{fig:chain_TL}].

Our results indicate that the diagonal temperature patterns observed in Fig. \ref{fig:180113a_Tis}(b) and in the classical molecular dynamics simulations of Ref. \cite{saaskilahti12} may be washed out by the quantum effects at low temperature. At higher temperature, quantum effects are reduced and the diagonal features reappear, but only if phonon transport remains close to ballistic. Increasing the temperature also increases phonon-phonon scattering \cite{ziman}, so finding a temperature regime where classical statistics prevail but phonon transport is sufficiently ballistic can be problematic.

The self-consistently modeled center region of Fig. \ref{fig:180113a_Tis} contains 4873 atoms. For this size of system and temperature range, the determination of the quantum linear response temperature profile, which was used as the initial guess for Newton-Raphson iteration, took approximately five hours wall-time with 12 CPU cores. Newton-Raphson iteration converged after three iterations and took approximately 14 hours wall-time. The calculation of the classical temperature profile took approximately 14 hours wall-time as well. The solution of the classical temperature profile is computationally more demanding than calculating the linear response profile, since the population functions appearing in the equations decay more slowly and need more integration time. 

Note that even though the system is smaller than the mean free path of long-wavelength phonons, the use of non-reflecting boundary conditions (i.e., semi-infinite leads) ensures that phonons are not reflected from the boundaries between the center region and the leads back to the junction. If the ends had been thermalized with Ohmic heat baths, reflections from the baths could skew the temperature profiles.

\subsection{Thermal transport in a graphene constriction}
\label{sec:graphene}
\begin{figure}
 \begin{center}
 \includegraphics[width=8.0cm]{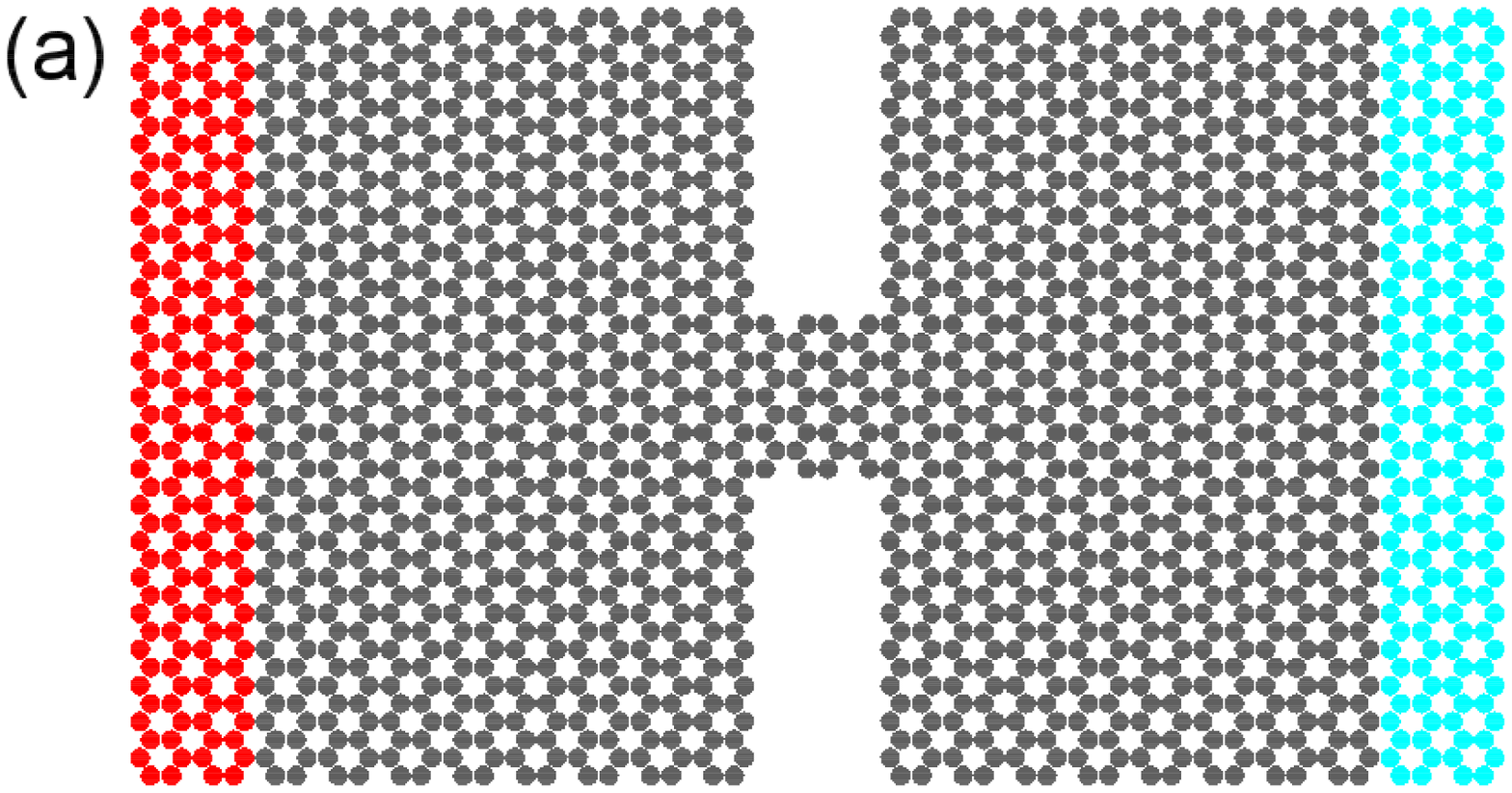}
 \includegraphics[width=8.6cm]{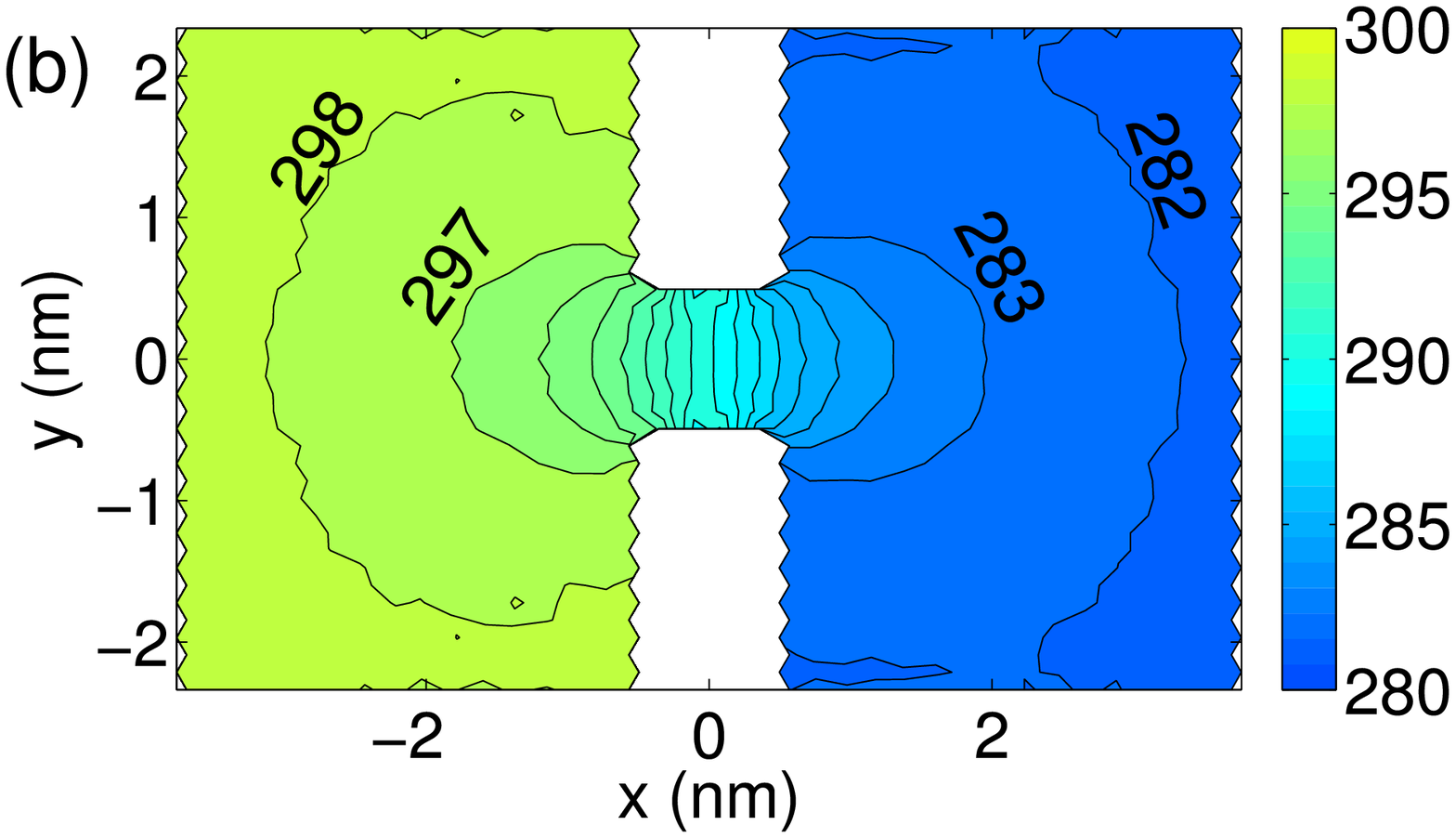}
 \includegraphics[width=8.6cm]{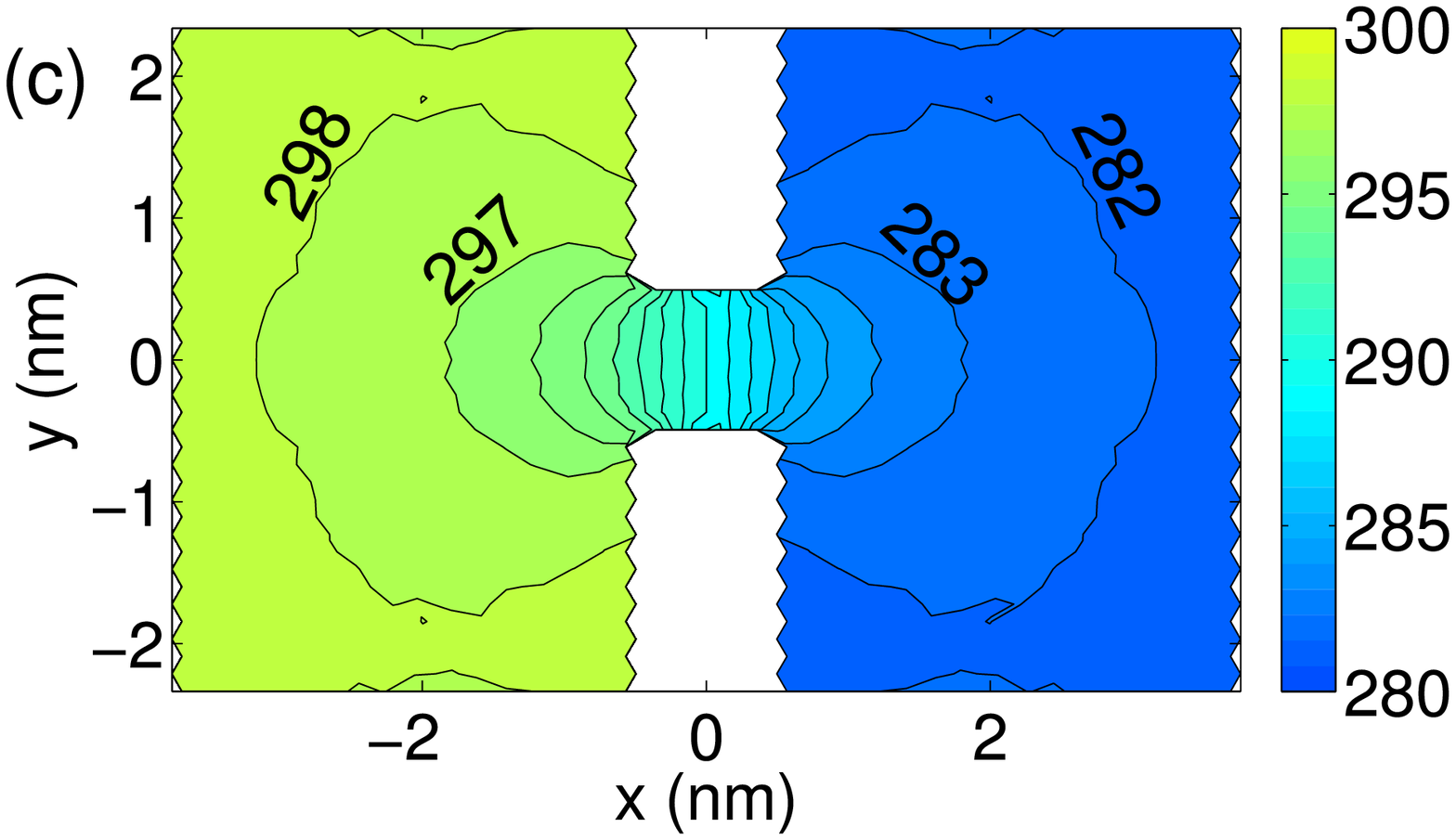}
 \end{center}
 \caption{(Color online) (a) Graphene nanoconstriction. The leads extend infinitely to the left and right, but the temperatures are determined self-consistently only for the gray atoms in the shown center region. (b) and (c) Self-consistent bath temperature profiles (K), in (b) quantum exact case and (c) classical approximation. The semi-infinite leads are at temperatures $T_L=300$ K and $T_R=280$ K. The relaxation time $\tau=1/\gamma$ is set to $1$ ps.}
 \label{fig:190113a}
\end{figure}

The two-dimensional square lattice model is easily extended to real materials such as graphene. It is interesting to see, for example, if the directional features observed in the square lattice remain for more complex lattice geometries. The example geometry is shown in Fig. \ref{fig:190113a}(a). The junction geometry in graphene has also been studied earlier \cite{xu10}, but our methodology gives access to local temperature profiles in the constriction. The method also allows to include diffusive effects, which would become important in large systems where the mean-free path is comparable to device dimensions. Each atom now has three degrees of freedom, which are all coupled to a single local Langevin heat bath. We set the temperature range close to the room temperature, $T_L=300$ K and $T_R=280$ K, because the acoustic phonon lifetime $\tau$ at room temperature is known to be of the order of $\tau=1$ ps \cite{bonini12}, suggesting that the bath coupling constant is $\gamma=\tau^{-1}=10^{12}$ s$^{-1}$. Carbon-carbon interactions are modeled by the fourth-nearest-neighbor force constant model \cite{saito} with the parameters of Ref. \cite{wirtz04}, which reproduce the bulk ab-initio phonon spectrum of graphene very accurately, at least for the acoustic modes. The optical modes are not active at room temperature, since they are populated only at temperatures close to $T\approx 1000$ K, but are fully included in the model in any case. We have verified the correct implementation of the force constant model by comparing ballistic thermal conductances of pure nanoribbons to the results of Ref. \cite{huang10}.  

Figures \ref{fig:190113a}(b) and \ref{fig:190113a}(c) show the quantum and classical temperature profiles close to the room temperature. The temperature profiles agree quite closely, which is unexpected since the phonon populations originating from the classical and Bose-Einstein distributions at room temperature are quite different: the highest-lying vibrational energies of graphene correspond to temperatures of $T_D\approx 2300 $ K. The agreement of temperature profiles therefore suggests that only the low-frequency modes close to the $\Gamma$ point, for which quantum and classical statistics agree, contribute to the transport and detailed temperature profile. On the other hand, the heat flow through the structure is still quite strongly overestimated by the classical approximation: quantum current is $Q\approx 2.1\times 10^{-8} $ W and the classical $Q \approx 5.1\times 10^{-8}$ W. No directional features appear in the studied geometry at room temperature, but lowering the temperature and increasing the phonon relaxation time could produce more complex temperature profiles. These studies, as well as investigation of different geometries and their influence on temperature profiles, are left for future work. Note also that approximately only half of the total temperature drop takes place in the constriction. 

\subsection{Comparison of the solution methods}
\label{sec:results_comparison}

As a final example of the numerics of the solutions, we compare the Newton-Raphson iteration and the ordinary differential equation (ODE) method. The differential equation \eqref{eq:dTdt} is integrated using the MATLAB\textsuperscript{\textregistered} \cite{matlab12b} implementation of an explicit Runge-Kutta formula with the Dormand-Prince pair \cite{dormand80} and adaptive step-size. Using an adaptive step-size integrator is necessary to avoid slowing down as integration approaches the self-consistent temperature configuration. Integration is stopped when the maximum heat current flowing to the bath is less than $10^{-5}\gamma$. 

\begin{figure}
 \includegraphics[width=8.6cm]{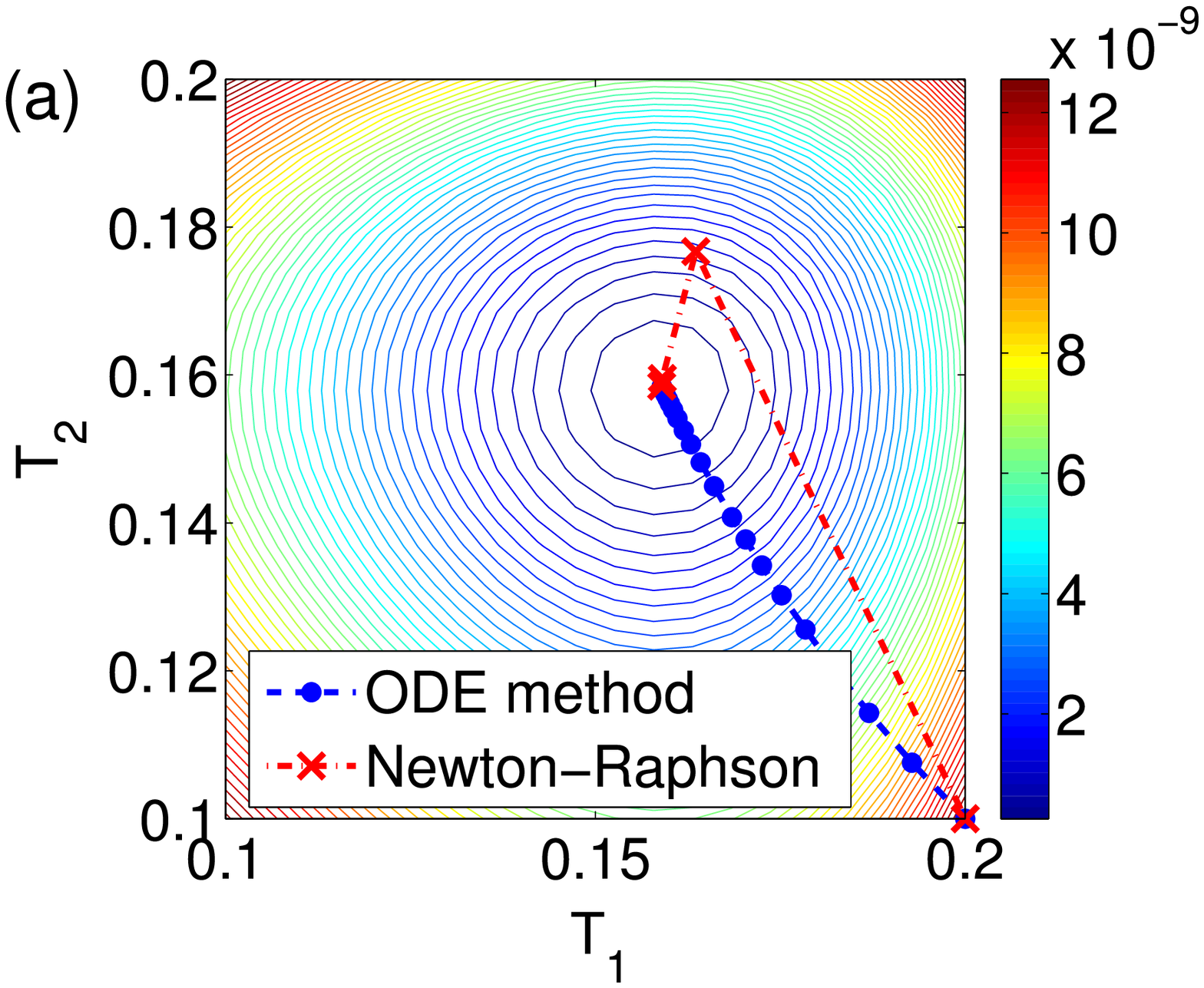}
 \includegraphics[width=8.6cm]{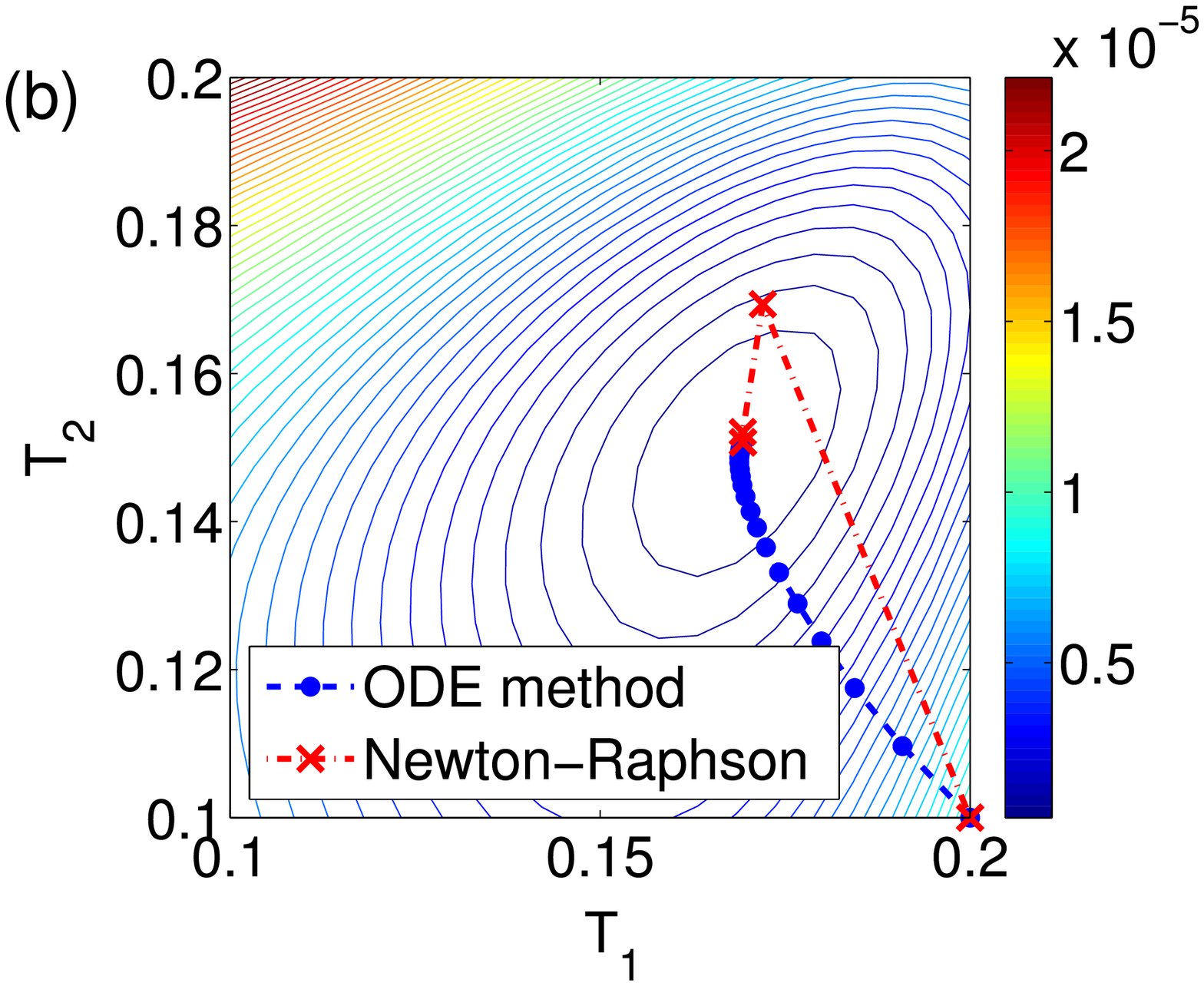}
 \caption{(Color online) The search for the self-consistent bath temperatures $T_1$ and $T_2$ that satisfy $Q_1(T_1,T_2)=Q_2(T_1,T_2)=0$ for an $N=2$ chain. The lead temperatures are $T_L=0.2$ and $T_R=0.1$ and the friction parameters are (a) $\gamma=10^{-2}$ and (b) $\gamma=1$. The two methods used for the search are Runge-Kutta-Fehlberg integration of Eq. \eqref{eq:dTdt} (dots connected by dashed lines) and Newton-Raphson iteration (crosses connected by dash-dotted lines). The contours belong to the ''target function'' $f(T_1,T_2)$ that is defined to be the squared sum of the currents flowing to the self-consistent reservoirs, $f(T_1,T_2)=Q_1(T_1,T_2)^2+Q_2(T_1,T_2)^2$. The self-consistent temperature configuration $(\tilde T_1,\tilde T_2)$ satisfies $f(\tilde T_1,\tilde T_2)=0$ and is also the global minimum of $f$. The initial guess is $T_1=0.2$, $T_2=0.1$.}
 \label{fig:comparison_rkf_nr}
\end{figure}

Figure \ref{fig:comparison_rkf_nr} shows the comparison of the two methods in the search for self-consistent solution. The setup is the Rubin-Greer setup of Fig. \ref{fig:chain_rg}(a) with $N=2$ and the values of the friction parameter are (a) $\gamma=0.01$, and (b) $\gamma=1$. Although it is best to use, e.g., the linear response approximation temperatures as initial guess for iteration, we use now $T_1=T_L=0.2$ and $T_2=T_R=0.1$ for illustrative purposes. The contour lines of the target function $f(T_1,T_2)=Q_1^2+Q_2^2$ are also shown. The function $f$ is defined such that the self-consistent temperature configuration is the global minimum and zero of $f$. The self-consistent temperatures are $(\tilde T_1,\tilde T_2)=(0.1590,0.1585)$ and $(\tilde T_1,\tilde T_2)=(0.1616,0.1574)$ for the cases of Figs. \ref{fig:comparison_rkf_nr}(a) and \ref{fig:comparison_rkf_nr}(b), respectively. 

For both weak and strong friction, the Newton-Raphson iteration proceeds similarly: The first iteration step of the Newton-Raphson method slightly misses the solution, but the second iteration already takes temperatures very close to the self-consistent temperature configuration. ODE method, on the other hand, proceeds approximately along the direction of steepest descent in the target function. Since the gradient $\nabla f = 2 \bb{J}^T \bb{Q}$, $\bb{J}$ being the Jacobian matrix of $\bb{Q}$, is not necessarily parallel to $\bb{Q}$, the path taken by the ODE method is generally not strictly along the steepest descent. 

For the case of larger $\gamma$ in Fig. \ref{fig:comparison_rkf_nr}(b), the contour lines are elongated forming a canyon-like shape and the heat exchange between the local baths starts to dominate over the heat exchange with the leads. In this case, ODE method does not proceed directly towards the solution. We have noted that such cases can be very difficult to handle for the ODE method, since the residual time integration along the canyon requires a very small step size. Newton-Raphson iteration, on the other hand, always seems to find the solution with only a few iterations.

In future, it would be interesting to study how well phonon damping and dephasing induced by the self-consistent heat baths mimics true anharmonic effects. Comparisons could be carried out, for example, by comparing the classical approximation of self-consistent equations with classical molecular dynamics (MD) simulations. Knowing now that the bath temperatures correspond to the local kinetic temperature [Eq. \eqref{eq:ekin_classical}], the local bath temperatures could be meaningfully compared to the local kinetic energy densities obtained from MD. It would also be worth investigating whether non-diffusive transport effects such as anomalous heat conduction in one dimension \cite{dhar08} could be reproduced by using a frequency-dependent bath coupling constant. If that is the case, one could study quantum effects in anomalous transport using the Green's function method.

\section{Conclusions}
\label{sec:conclusions}

We studied quantum heat transport in nanostructures using the Green's function method combined with self-consistent heat baths. Semi-infinite leads acting as thermal reservoirs were reduced to sources of noise and dissipation in the boundaries of the scattering region. In the scattering region, the temperatures of the heat baths mimicking anharmonic effects were determined self-consistently from the requirement of heat current conservation. The self-consistent bath temperatures were shown to measure local energy density, thereby giving them a meaningful physical interpretation. In the classical limit, local kinetic temperature is equal to the bath temperature. 

By coupling one-dimensional chain to semi-infinite chains, thereby eliminating contact resistance, we demonstrated the equivalence of thermal currents obtained by the self-consistent thermal bath model and the Boltzmann transport equation under gray approximation with full phonon dispersion. Self-consistent thermal bath model is, therefore, a physically meaningful method to introduce phonon relaxation to ballistic quantum transport models.

As an application of the formalism, we presented temperature profiles in two-dimensional constrictions and showed that quantum statistics plays a vital role in how directional patterns of temperature emanate from the junction. In a graphene constriction at room temperature, the bath temperature profile obtained by classical approximation agreed very closely with the quantum temperature profile, suggesting that quantum effects are not strong and molecular dynamics simulations could be justified under those assumptions. In more general cases, we expect, however, that the Green's function method combined with self-consistent thermal baths is a very useful tool in studying quantum heat transfer in the ballistic, diffusive and crossover regimes of phonon transport due to the good balance of complexity, insight and predictivity it offers.

\section{Acknowledgements}
We acknowledge the Finnish IT Center for Science and the Aalto Science-IT project for computer time. The work is in part funded by the MIDE and AEF research programs of Aalto University and by the Graduate School in Electronics, Telecommunications and Automation (GETA). 

\appendix*
 \section{Derivation of Eq. \eqref{eq:qlandauer_1}}
 \label{sec:appendix}
In this appendix, we derive Eq. \eqref{eq:qlandauer_1} for thermal current flowing to a local heat bath. For notational simplicity, we combine the lead noises $\eta_L$ and $\eta_R$ and center region local bath noises $\xi_{C}$ to a single vector variable $\bb{s}^J$, where the index $J\in\{L,R,1,2,\dots,N_C\}$ is now a general index for either a lead bath ($J\in\{L,R\}$) or a self-consistent local bath ($J\in\{1,2,\dots,N_C\}$). $N_C$ is the number of atoms in the center region. Explicitly, $\bb{s}^L=\eta_L$, $\bb{s}^R=\eta_R$ and $\bb{s}^i$ is a vector whose only non-zero component is $s^i_i=\xi_{Ci}$, the $i$th component of $\xi_C$. The noise covariances are then
\begin{equation}
 \langle \hat{\bb{s}}^J(\omega) \hat{\bb{s}}^{J'}(\omega') \rangle = 2\pi \delta(\omega+\omega') \Gamma^J(\omega) \left[ f_B(\omega,T_J) +1 \right] \delta^{JJ'}.
\end{equation}
If index $J\in\{L,R\}$, the coupling function $\Gamma^J$ is defined by the self-energy of the lead, Eq. \eqref{eq:gammaIdef}. For $J=i$, the only non-zero matrix element in the coupling function is $\Gamma_{ii}^i(\omega) = 2m\gamma\omega$ (we write $\gamma_C=\gamma$ in this section).

Equation \eqref{eq:sol_uom} can be written (dropping the index $C$ for center region)
\begin{equation}
 \hat{\bb{u}}(\omega) = -\bb{G}(\omega) \sum_J \hat{\bb{s}}^J(\omega).
\end{equation}
The heat flowing to an Ohmic bath is obtained by calculating the statistical average of the symmetrized heat current
\begin{equation}
 Q_i = \gamma m \dot{u}_i^2 - \frac{1}{2} [\dot{u}_i \xi_i + \xi_i \dot{u}_i].
\end{equation}
We proceed term by term. The statistical average of the first term is
\begin{alignat}{2}
 \langle & \gamma m \dot{u}_i^2 \rangle \notag \\
 &= \gamma m \left\langle \int \frac{d\omega}{2\pi} (-i\omega)\hat{u}_i(\omega) e^{-i\omega t} \int \frac{d\omega'}{2\pi} (-i\omega')\hat{u}_i(\omega') e^{-i\omega' t} \right\rangle \\
  &=  \int \frac{d\omega}{2\pi} \frac{d\omega'}{2\pi}\gamma m  (-\omega\omega') e^{-i(\omega+\omega') t} \sum_{JJ'} \sum_{jk} G_{ij}(\omega) G_{ik}(\omega') \notag \\
  &\quad \times \left\langle \hat{s}^J_j(\omega)  \hat{s}^{J'}_k(\omega') \right\rangle  \\
  &=   \int \frac{d\omega}{2\pi} \gamma m  \omega^2 \sum_J \left[\bb{G}(\omega) \Gamma^J(\omega) \bb{G}(-\omega)^T \right]_{ii} \left[f_B(\omega,T_J)+1\right]  .
 \label{eq:qi_firstterm}
\end{alignat}
The average of the second term is
\begin{alignat}{2}
 - \langle \dot{u}_i \xi_i \rangle &= - \left\langle \int \frac{d\omega}{2\pi} (-i\omega) \hat{u}_i(\omega) e^{-i\omega t} \int \frac{d\omega'}{2\pi} \hat{\xi}_i(\omega') e^{-i\omega' t} \right\rangle  \\
  &= \int \frac{d\omega}{2\pi} \frac{d\omega'}{2\pi}(-i\omega) e^{-i(\omega+\omega') t} \sum_J \sum_{j} G_{ij}(\omega) \notag \\
  &\quad \times \left\langle \hat{s}^J_j(\omega) \hat{\xi}_i(\omega') \right\rangle .
\end{alignat}
The only term surviving the sum over baths $J$ is the one corresponding to the local heat bath at site $i$, so
\begin{equation}
 -\langle \dot{u}_i \xi_i \rangle = - 2 \int \frac{d\omega}{2\pi} i G_{ii}(\omega) \gamma m \omega^2 \left[f_B(\omega,T_i)+1\right].
\end{equation}
Combined with the symmetrizing term, one gets 
\begin{alignat}{2}
 - \frac{1}{2} \langle \dot{u}_i \xi_i + \xi_i \dot{u}_i \rangle &= - \int \frac{d\omega}{2\pi} i[ G_{ii}(\omega)-G_{ii}(-\omega)] \gamma m \omega^ 2 \notag \\
  &\quad \times \left[f_B(\omega,T_i)+1\right] \label{eq:a10} \\
 &= -  \int \frac{d\omega}{2\pi}  \sum_J [\bb{G}(\omega) \Gamma^J(\omega) \bb{G}(-\omega)]_{ii} \notag \\
  &\quad \times \gamma m \omega^2 \left[f_B(\omega,T_i)+1\right],\label{eq:qi_secondterm} 
\end{alignat}
where we used Eq. \eqref{eq:gminusg} with the replacements $\bb{g}_I\to \bb{G}$ and $\tilde \Gamma^I\to \sum_J\Gamma^J$. Combining Eqs. \eqref{eq:qi_firstterm} and \eqref{eq:qi_secondterm}, we get
\begin{alignat}{2}
 \langle Q_i \rangle &= \int \frac{d\omega}{2\pi} \gamma m \omega^2 \sum_J \left[\bb{G}(\omega) \Gamma^J(\omega) \bb{G}(-\omega)^T \right]_{ii} \notag \\
  &\quad \times \left[f_B(\omega,T_J)-f_B(\omega,T_i)\right]  .
\end{alignat}
Noting that the integrand is an even function finally gives Eq. \eqref{eq:qlandauer_1}.


\begin{thebibliography}{10}

\bibitem{berber00}
S.~Berber, Y.-K. Kwon, and D.~Tom\'anek,
\newblock Phys. Rev. Lett. {\bf 84}, 4613 (2000).

\bibitem{kim01}
P.~Kim, L.~Shi, A.~Majumdar, and P.~L. McEuen,
\newblock Phys. Rev. Lett. {\bf 87}, 215502 (2001).

\bibitem{chang08}
C.~W. Chang, D.~Okawa, H.~Garcia, A.~Majumdar, and A.~Zettl,
\newblock Phys. Rev. Lett. {\bf 101}, 075903 (2008).

\bibitem{rego98}
L.~G.~C. Rego and G.~Kirczenow,
\newblock Phys. Rev. Lett. {\bf 81}, 232 (1998).

\bibitem{schwab00}
K.~Schwab, E.~A. Henriksen, J.~M. Worlock, and M.~L. Roukes,
\newblock Nature {\bf 404}, 974 (2000).

\bibitem{prunnila10}
M.~Prunnila and J.~Meltaus,
\newblock Phys. Rev. Lett. {\bf 105}, 125501 (2010).

\bibitem{altfeder10}
I.~Altfeder, A.~A. Voevodin, and A.~K. Roy,
\newblock Phys. Rev. Lett. {\bf 105}, 166101 (2010).

\bibitem{majumdar04}
A.~Majumdar,
\newblock Science {\bf 303}, 777 (2004).

\bibitem{dubi11}
Y.~Dubi and M.~Di~Ventra,
\newblock Rev. Mod. Phys. {\bf 83}, 131 (2011).

\bibitem{pop10}
E.~Pop,
\newblock Nano Res. {\bf 3}, 147 (2010).

\bibitem{terraneo02}
M.~Terraneo, M.~Peyrard, and G.~Casati,
\newblock Phys. Rev. Lett. {\bf 88}, 094302 (2002).

\bibitem{li06}
B.~Li, L.~Wang, and G.~Casati,
\newblock Appl. Phys. Lett. {\bf 88}, 143501 (2006).

\bibitem{chang06}
C.~W. Chang, D.~Okawa, A.~Majumdar, and A.~Zettl,
\newblock Science {\bf 314}, 1121 (2006).

\bibitem{murthy05}
J.~Y. Murthy {\em et~al.},
\newblock Int. J. Multiscale Comp. Eng. {\bf 3}, 5 (2005).

\bibitem{landauer70}
R.~Landauer,
\newblock Philos. Mag. {\bf 21}, 863 (1970).

\bibitem{buttiker92}
M.~B\"uttiker,
\newblock Phys. Rev. B {\bf 46}, 12485 (1992).

\bibitem{mingo03}
N.~Mingo and L.~Yang,
\newblock Phys. Rev. B {\bf 68}, 245406 (2003).

\bibitem{mingo06}
N.~Mingo,
\newblock Phys. Rev. B {\bf 74}, 125402 (2006).

\bibitem{wang06}
J.-S. Wang, J.~Wang, and N.~Zeng,
\newblock Phys. Rev. B {\bf 74}, 033408 (2006).

\bibitem{wang08}
J.-S. Wang, J.Wang, and J.~L\"u,
\newblock Eur. Phys. J. B {\bf 62}, 381 (2008).

\bibitem{bolsterli70}
M.~Bolsterli, M.~Rich, and W.~M. Visscher,
\newblock Phys. Rev. A {\bf 1}, 1086 (1970).

\bibitem{bonetto04}
F.~Bonetto, J.~L. Lebowitz, and J.~Lukkarinen,
\newblock J. Stat. Phys. {\bf 116}, 783 (2004).

\bibitem{pereira04}
E.~Pereira and R.~Falcao,
\newblock Phys. Rev. E {\bf 70}, 046105 (2004).

\bibitem{dhar06}
A.~Dhar and D.~Roy,
\newblock J. Stat. Phys. {\bf 125}, 801 (2006).

\bibitem{roy08}
D.~Roy,
\newblock Phys. Rev. E {\bf 77}, 062102 (2008).

\bibitem{visscher75}
W.~M. Visscher and M.~Rich,
\newblock Phys. Rev. A {\bf 12}, 675 (1975).

\bibitem{bandyopadhyay11}
M.~Bandyopadhyay and D.~Segal,
\newblock Phys. Rev. E {\bf 84}, 011151 (2011).

\bibitem{falcao08}
R.~Falcao, A.~Neto, and E.~Pereira,
\newblock Theoretical and Mathematical Physics {\bf 156}, 1081 (2008).

\bibitem{bonetto09}
F.~Bonetto, J.~Lebowitz, J.~Lukkarinen, and S.~Olla,
\newblock J. Stat. Phys. {\bf 134}, 1097 (2009).

\bibitem{pereira13}
E.~Pereira, R.~Falcao, and H.~C.~F. Lemos,
\newblock Phys. Rev. E {\bf 87}, 032158 (2013).

\bibitem{barros06}
F.~Barros, H.~C.~F. Lemos, and E.~Pereira,
\newblock Phys. Rev. E {\bf 74}, 052102 (2006).

\bibitem{neto07}
A.~F. Neto, H.~C.~F. Lemos, and E.~Pereira,
\newblock Phys. Rev. E {\bf 76}, 031116 (2007).

\bibitem{santana12}
L.~M. Santana and E.~Pereira,
\newblock Phys. Rev. E {\bf 86}, 032105 (2012).

\bibitem{pereira08}
E.~Pereira and H.~C.~F. Lemos,
\newblock Phys. Rev. E {\bf 78}, 031108 (2008).

\bibitem{segal09}
D.~Segal,
\newblock Phys. Rev. E {\bf 79}, 012103 (2009).

\bibitem{pereira10}
E.~Pereira, 
\newblock Phys. Lett. A {\bf{374}}, 1933 (2010).

\bibitem{pereira11}
E.~Pereira, H.~C.~F. Lemos, and R.~R. \'Avila,
\newblock Phys. Rev. E {\bf 84}, 061135 (2011).

\bibitem{avila13}
R.~R. Ávila and E.~Pereira,
\newblock J. Phys. A: Math. and Theor. {\bf 46}, 055002 (2013).

\bibitem{buttiker85}
M.~B\"uttiker,
\newblock Phys. Rev. B {\bf 32}, 1846 (1985).

\bibitem{buttiker86}
M.~B\"uttiker,
\newblock Phys. Rev. B {\bf 33}, 3020 (1986).

\bibitem{damato90}
J.~L. D'Amato and H.~M. Pastawski,
\newblock Phys. Rev. B {\bf 41}, 7411 (1990).

\bibitem{roy07}
D.~Roy and A.~Dhar,
\newblock Phys. Rev. B {\bf 75}, 195110 (2007).

\bibitem{golizadeh07}
R.~Golizadeh-Mojarad and S.~Datta,
\newblock Phys. Rev. B {\bf 75}, 081301 (2007).

\bibitem{dhar03}
A.~Dhar and B.~S. Shastry,
\newblock Phys. Rev. B {\bf 67}, 195405 (2003).

\bibitem{rubin71}
R.~J. Rubin and W.~L. Greer,
\newblock J. Math. Phys. {\bf 12}, 1686 (1971).

\bibitem{ford65}
G.~W. Ford, M.~Kac, and P.~Mazur,
\newblock J. Math. Phys. {\bf 6}, 504 (1965).

\bibitem{weiss}
U.~Weiss,
\newblock {\em Quantum Dissipative Systems}, 3rd ed. (World Scientific,
  Singapore, 2008).

\bibitem{ziman}
J.~Ziman,
\newblock {\em Electrons and Phonons: The Theory of Transport Phenomena in
  Solids} (Oxford University Press, USA, 2001).

\bibitem{ford88}
G.~W. Ford, J.~T. Lewis, and R.~F. O'Connell,
\newblock Phys. Rev. A {\bf 37}, 4419 (1988).

\bibitem{lopezsancho85}
M.~P. {Lopez Sancho}, J.~M. {Lopez Sancho}, J.~M.~L. Sancho, and J.~Rubio,
\newblock J. Phys. F: Met. Phys. {\bf 15}, 851 (1985).

\bibitem{li09jap}
N.~Li and B.~Li,
\newblock J. Phys. Soc. Jpn. {\bf 78}, 044001 (2009).

\bibitem{wang07}
J.-S. Wang,
\newblock Phys. Rev. Lett. {\bf 99}, 160601 (2007).

\bibitem{hardy63}
R.~J. Hardy,
\newblock Phys. Rev. {\bf 132}, 168 (1963).

\bibitem{caroli71}
C.~Caroli, R.~Combescot, P.~Nozieres, and D.~Saint-James,
\newblock J. Phys. C: Solid State Phys. {\bf 4}, 916 (1971).

\bibitem{yamamoto06}
T.~Yamamoto and K.~Watanabe,
\newblock Phys. Rev. Lett. {\bf 96}, 255503 (2006).

\bibitem{angelescu98}
D.~E. Angelescu, M.~C. Cross, and M.~L. Roukes,
\newblock Superlattices Microstruct. {\bf 23}, 673  (1998).

\bibitem{segal03}
D.~Segal, A.~Nitzan, and P.~Hänggi,
\newblock J. Chem. Phys. {\bf 119}, 6840 (2003).

\bibitem{zhang07}
W.~Zhang, T.~S. Fisher, and N.~Mingo,
\newblock Numer. Heat Transfer, Part B {\bf 51}, 333 (2007).

\bibitem{goldstein}
H.~Goldstein,
\newblock {\em Classical Mechanics} (Addison-Wesley Publishing Company, Inc.,
  1971).

\bibitem{mahan}
G.~D. Mahan,
\newblock {\em Many Particle Physics}, 3rd ed. (Kluwer Academic/Plenum
  Publishers, New York, 2010).

\bibitem{lepri03}
S.~Lepri, R.~Livi, and A.~Politi,
\newblock Phys. Rep. {\bf 377}, 1 (2003).

\bibitem{majumdar93}
A.~Majumdar,
\newblock J. Heat Transfer {\bf 115}, 7 (1993).

\bibitem{minnich11}
A.~J. Minnich, G.~Chen, S.~Mansoor, and B.~S. Yilbas,
\newblock Phys. Rev. B {\bf 84}, 235207 (2011).

\bibitem{hopkins09}
P.~E. Hopkins, P.~M. Norris, M.~S. Tsegaye, and A.~W. Ghosh,
\newblock Journal of Applied Physics {\bf 106}, 063503 (2009).

\bibitem{saha07}
S.~K. Saha and L.~Shi,
\newblock J. Appl. Phys. {\bf 101}, 074304 (2007).

\bibitem{saaskilahti12}
K.~S\"a\"askilahti, J.~Oksanen, R.~P. Linna, and J.~Tulkki,
\newblock Phys. Rev. E {\bf 86}, 031107 (2012).

\bibitem{bartsch12}
T.~Bartsch, M.~Schmidt, C.~Heyn, and W.~Hansen,
\newblock Phys. Rev. Lett. {\bf 108}, 075901 (2012).

\bibitem{xu10}
Y.~Xu, X.~Chen, J.-S. Wang, B.-L. Gu, and W.~Duan,
\newblock Phys. Rev. B {\bf 81}, 195425 (2010).

\bibitem{bonini12}
N.~Bonini, J.~Garg, and N.~Marzari,
\newblock Nano Lett. {\bf 12}, 2673 (2012).

\bibitem{saito}
R.~Saito, G.~Dresselhaus, and M.~S. Dresselhaus,
\newblock {\em Physical Properties Of Carbon Nanotubes} (Imperial College
  Press, 1998).

\bibitem{wirtz04}
L.~Wirtz and A.~Rubio,
\newblock Solid State Comm. {\bf 131}, 141  (2004).

\bibitem{huang10}
Z.~Huang, T.~S. Fisher, and J.~Y. Murthy,
\newblock J. Appl. Phys. {\bf 108}, 094319 (2010).

\bibitem{matlab12b}
MATLAB,
\newblock {\em 8.0.0.783 (R2012b)} (The MathWorks Inc., Natick, Massachusetts,
  2012).

\bibitem{dormand80}
J.~Dormand and P.~Prince,
\newblock J. Comp. Appl. Math. {\bf 6}, 19  (1980).

\bibitem{dhar08}
A.~Dhar,
\newblock Adv. Phys. {\bf 57}, 457 (2008).

\end{thebibliography}


\end{document}